\PassOptionsToPackage{hyphens}{url}
\PassOptionsToPackage{dvipsnames,svgnames,x11names}{xcolor}
\documentclass[
  12pt]{article}

\usepackage[table]{xcolor}
\usepackage{booktabs}
\usepackage{graphicx}
\usepackage{siunitx}
\usepackage{geometry}
\usepackage{algorithm}
\usepackage{algpseudocode}
\usepackage{natbib}
\definecolor{cbLow}{gray}{0.55}
\definecolor{cbHigh}{gray}{0.97}

\usepackage{amsmath,amssymb,mathrsfs,amsthm,bm}
\usepackage{nopageno}
\usepackage{verbatim}
\usepackage{multirow}
\usepackage{enumitem}
\setlist{topsep=0.3em, itemsep=0.15em, parsep=0pt}
\usepackage{bbm}
\usepackage{pdfpages}
\usepackage{tcolorbox}
\usepackage{cancel}
\usepackage{iftex}
\ifPDFTeX
  \usepackage[T1]{fontenc}
  \usepackage[utf8]{inputenc}
  \usepackage{textcomp} 
\else 
  \usepackage{unicode-math}
  \defaultfontfeatures{Scale=MatchLowercase}
  \defaultfontfeatures[\rmfamily]{Ligatures=TeX,Scale=1}
\fi
\usepackage{lmodern}
\ifPDFTeX\else
\fi
\IfFileExists{upquote.sty}{\usepackage{upquote}}{}
\IfFileExists{microtype.sty}{
  \usepackage[]{microtype}
  \UseMicrotypeSet[protrusion]{basicmath} 
}{}
\usepackage{titlesec}
\titlespacing*{\section}{0pt}{1.5ex plus .5ex minus .2ex}{0.3ex plus .1ex}
\titlespacing*{\subsection}{0pt}{0.3ex plus .1ex minus .1ex}{0.5ex plus .1ex}
\titlespacing*{\subsubsection}{0pt}{1.25ex plus .4ex minus .2ex}{0.5ex plus .2ex}
\titlespacing*{\paragraph}{0pt}{1ex plus .3ex minus .2ex}{1em}

\newcommand*{\parencite}{\citep}
\newcommand{\given}{\,|\,}

\usepackage{longtable,array}
\usepackage{calc} 
\usepackage{etoolbox}
\makeatletter
\patchcmd\longtable{\par}{\if@noskipsec\mbox{}\fi\par}{}{}
\makeatother
\IfFileExists{footnotehyper.sty}{\usepackage{footnotehyper}}{\usepackage{footnote}}
\makesavenoteenv{longtable}
\makeatletter
\def\maxwidth{\ifdim\Gin@nat@width>\linewidth\linewidth\else\Gin@nat@width\fi}
\def\maxheight{\ifdim\Gin@nat@height>\textheight\textheight\else\Gin@nat@height\fi}
\makeatother
\setkeys{Gin}{width=\maxwidth,height=\maxheight,keepaspectratio}
\makeatletter
\def\fps@figure{htbp}
\makeatother

\makeatletter
\@ifpackageloaded{caption}{}{\usepackage{caption}}
\AtBeginDocument{%
\ifdefined\contentsname
  \renewcommand*\contentsname{Table of contents}
\else
  \newcommand\contentsname{Table of contents}
\fi
\ifdefined\listfigurename
  \renewcommand*\listfigurename{List of Figures}
\else
  \newcommand\listfigurename{List of Figures}
\fi
\ifdefined\listtablename
  \renewcommand*\listtablename{List of Tables}
\else
  \newcommand\listtablename{List of Tables}
\fi
\ifdefined\figurename
  \renewcommand*\figurename{Figure}
\else
  \newcommand\figurename{Figure}
\fi
\ifdefined\tablename
  \renewcommand*\tablename{Table}
\else
  \newcommand\tablename{Table}
\fi
}
\@ifpackageloaded{float}{}{\usepackage{float}}
\floatstyle{ruled}
\@ifundefined{c@chapter}{\newfloat{codelisting}{h}{lop}}{\newfloat{codelisting}{h}{lop}[chapter]}
\floatname{codelisting}{Listing}

\makeatother
\makeatletter
\@ifpackageloaded{subcaption}{}{\usepackage{subcaption}}
\makeatother

\ifLuaTeX
  \usepackage{selnolig}  
\fi
\usepackage{bookmark}

\IfFileExists{xurl.sty}{\usepackage{xurl}}{} 
\hypersetup{
  pdftitle={Title},
  pdfauthor={Author 1; Author 2},
  pdfkeywords={3 to 6 keywords, that do not appear in the title},
  colorlinks=true,
  linkcolor={blue},
  filecolor={Maroon},
  citecolor={Blue},
  urlcolor={Blue},
  pdfcreator={LaTeX via pandoc}}

\newcommand{\anon}{1}

\begin{document}

\def\spacingset#1{\renewcommand{\baselinestretch}%
{#1}\small\normalsize} \spacingset{1}


\if1\anon
{
  \title{\bf Multilevel regression trees with application to wildfires in the American west}
  \author{John Henry V. Gray\\
  	Department of Statistics, Colorado State University
  	\\ \hspace{.2cm}\\
  	Tianjian Zhou \\ 
  	Takeda Pharmaceuticals \hspace{.2cm}\\
  	and \\ \hspace{.2cm}\\ 
  	Benjamin A. Shaby \\ 
  	Department of Statistics, Colorado State University}
  	
  \maketitle
} \fi

\if0\anon
{
  \bigskip
  \bigskip
  \bigskip
  \begin{center}
    {\LARGE\bf Title}
\end{center}
  \medskip
} \fi

\bigskip
\begin{abstract}
We propose a Bayesian regression tree model fit within a multilevel structure and apply it to historic wildfire data in the western United States. Sharing of information between related groups (ecoregions) combined with highly interpretable regression trees allows for better predictions and understanding of climate and land cover variables predictive of wildfires. By doing a simulation study with a range of performance metrics, we demonstrate our method produces tree posteriors most structurally similar to assumed true trees, while simultaneously achieving good out-of-sample predictive performance. Applied to a large wildfire data set, we explore variable splits within regression trees corresponding to each ecoregion in detail, taking into account known features of each location. Shared hyperparameters between trees provide highly useful understanding of both variable and split value importance in predicting wildfires among all ecoregions, with no direct parallel in comparable models. Namely, we highlight potential evaporation, temperature, and evergreen forest land cover as variables most associated with historic wildfires, with some observable patterns in split values most commonly chosen across the groups. We propose a new algorithm based on parallel tempering, conditioning on shared hyperparameters at the true posterior temperature, improving Markov chain mixing, a known bottleneck in Bayesian CART models.
\end{abstract}

\noindent%
\vfill

\newpage
\spacingset{1.8} 
\setlength{\abovedisplayskip}{8pt plus 2pt minus 4pt}
\setlength{\belowdisplayskip}{8pt plus 2pt minus 4pt}
\setlength{\abovedisplayshortskip}{4pt plus 2pt}
\setlength{\belowdisplayshortskip}{4pt plus 2pt}

\section{Introduction}\label{sec-intro}

In this work, we introduce a Bayesian multilevel regression tree model that enables sharing of information between ecoregions when predicting historic wildfires in the western United States. Our model produces valuable summaries of important factors predictive of wildfire size which are unattainable through other methods. At the same time, we retain all the benefits of individual trees, in contrast to ensemble methods, primarily ease of interpretation. We also introduce an efficient parallel tempering algorithm for multilevel models to enhance mixing, which has traditionally been a bottleneck for Bayesian tree models.

Wildfires in the United States continue to be a present and growing threat, with losses and response expenditure extending into billions of dollars \parencite{wildfire_review}. Understanding the environmental and climate factors associated with wildfire risk is an important part of mitigation efforts. The western US is home to regions particularly prone to wildfires. This part of the country is also comprised of unique environments and ecosystems not found elsewhere in the US. Consequently, wildfire occurrence here is worth careful study in its own right, separate from other areas of the country.

Classification and Regression Trees (CART) \parencite{Breiman_CART} form a class of statistical models with powerful predictive and explanatory capability. In their basic form, CART models partition data with a series of rectangular splits on predictor variables. They are among the most interpretable of models, with predictions easily traceable through a clear flow chart of splitting decisions.  Trees are also good at capturing complex interactions between variables that are harder to account for in traditional regression models \parencite{Elith_trees}.
Initially introduced as an algorithmic method, seminal work by \cite{Chipman98} and \cite{Denison98} expanded CART to a Bayesian framework, with enhanced understanding and propagation of uncertainty. Subsequent research considerably improved predictive performance  of tree-based models through ensemble methods such as Random Forests \parencite{Breiman2001} and Bayesian Additive Regression Trees (BART) \parencite{BART_chipman}. However, while measures of variable importance are available in ensemble methods, the use of a large number of trees to make predictions comes at a cost of reduced model interpretability. 

We are interested in understanding the factors most predictive of and associated with wildfires in the western US. We could apply any number of statistical and machine learning techniques like BART to historic wildfire data, retaining the advantages of tree-based methods to achieve good predictive performance at forecasting. However, many of these models would not be as useful at improving our understanding of these predictive factors. 

\begin{figure}[!ht]
    \centering
    \includegraphics[width=0.8\linewidth]{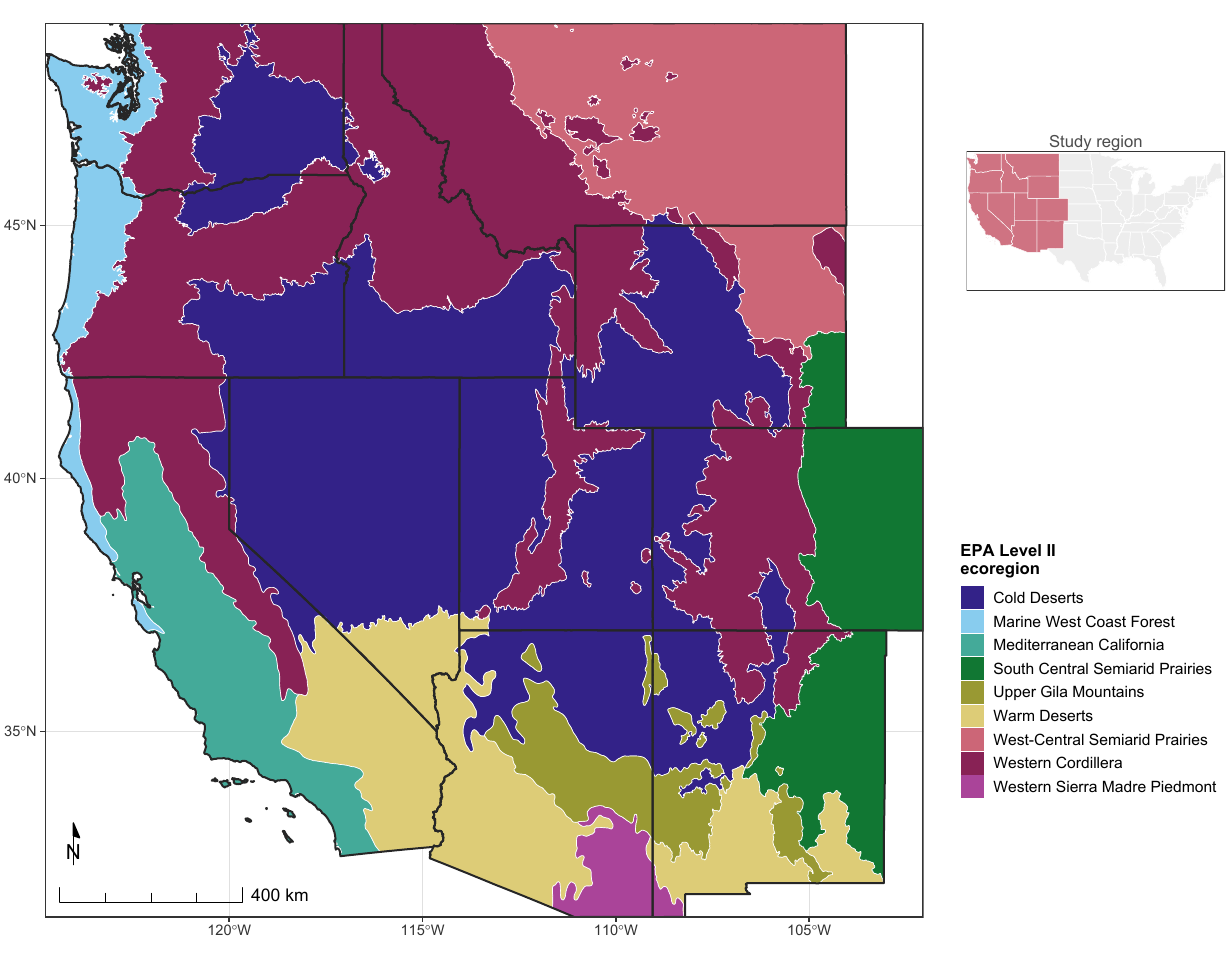}
    \caption{Map of level II ecoregions in the West. Source: \cite{EPA2010NACECEcoLevel2}.}
    \label{fig:ecoregions}
\end{figure}

The Commission for Environmental Cooperation in conjunction with the US Environmental Protection Agency develops geographical classifications for North America called Ecoregions \parencite{CEC1997,OmernikGriffith2014}, grouping areas with similar ecosystem, climate, and geographical characteristics. Ecoregions follow a hierarchy with four levels of increasing degrees of granularity. Level II of the hierarchy, which in the American West (defined here as the states of Washington, Oregon, California, Nevada, Idaho, Montana, Wyoming, Utah, Arizona, Colorado, and New Mexico) corresponds to 9 different ecoregions, is shown in Figure \ref{fig:ecoregions}. Breaking down historical wildfire data by ecoregions for analysis is a natural way to understand and predict wildfires. Namely, we expect that wildfires in different ecoregions share broadly similar predictive and explanatory factors related to climate, land cover, and other variables. At the same time, ecoregions also retain unique differences according to their different ecosystems. We will be exploiting these relationships between ecoregions in understanding and predicting wildfires in the West.

Our Bayesian multilevel regression tree model uses a collection of coupled individual regression trees fit to data with a natural grouping: wildfires broken down by ecoregions. Each tree is specific to its corresponding  ecoregion, with the growth of each tree informed by the structure of trees fitted in other ecoregions. This allows good variable splits to be shared among models within parts of the country with considerable overlap in wildfire occurrence, intensity, and likely similar associated climate and environmental contributing factors. These trees are fit within the context of a `partial pooling' model, where inference on distinct tree posteriors is done within each group, while trees are pooled towards a common structure for all data. The extent of pooling is determined by the amount of data within each ecoregion, with groups with less data able to borrow more from the combined, pooled data set. 

Tree-based approaches to multilevel data have typically focused on ensemble methods. \cite{Linero18} introduced a Dirichlet prior within BART to share predictor splitting information, termed Dirichlet Additive Regression Trees (DART). \cite{DuLineroOG_BART} built on DART, accounting for additional grouping structure among variables, termed the Overlapping Group Dirichlet prior. \cite{Goedhart_EB} employed an Empirical Bayes approach to estimating variable splitting probabilities supplemented with external covariate data termed EB-coBART, but again applied to BART models. \cite{jointBART} introduced jointBART, sharing splitting probabilities among groups in a network, involving a combination of Dirichlet and Markov Random Field priors. Other tree-based hierarchical models have focused on group-specific terms within BART terminal nodes  \parencite{wundervald2023hierarchicalembeddedbayesianadditive, prevot2025hierarchicalmodellingapproachbayesian}. To our knowledge, no existing work incorporates sharing of information via variable splitting probabilities among coupled, single regression trees in a multilevel setting.

By using Bayesian regression trees, we make use of the benefit of tree-based models with uncertainty quantification. Importantly, the use of single regression trees rather than ensemble tree methods facilitates excellent interpretability of predictions. Our method provides unique summaries of shared hyperparameters between ecoregions that are directly interpretable, with no counterpart in individual regression trees, as detailed in Section \ref{subsec:s_q_interpretation}. This enhanced interpretability enables us to achieve greater understanding of wildfire risks within a vast set of related areas of the US. We also introduce a new way of improving mixing within MCMC: a parallel tempering-within-Gibbs approach that fixes shared hyperparameters at the true posterior chain's values. All other chains are updated conditional upon these values.

\section{Multilevel regression tree model}

\paragraph*{Review of Bayesian regression trees}
\cite{Chipman98} and \cite{Denison98} independently introduced Bayesian methods for CART models. We follow the more popular approach introduced by \cite{Chipman98}. As the wildfire data response variable is quantitative, we will focus exclusively on regression trees. Let $\bm{X}$ and $\bm{Y}$ denote the matrix of predictors and response vector, respectively. A tree $T$ is defined as a collection of nodes and splits, with nodes being either internal or terminal. $T$ has associated parameters $\theta$. We assume a normal distribution with mean $\mu_b$ but common variance $\sigma^2$ in each terminal node, $b=1,\dots,B$ (mean shift model). The joint prior $p(\theta,T)$ is a product of the parameter prior $p(\theta\given T)$ and tree prior $P(T)$: $p(\theta,T) = p(\theta\given T)P(T).$ The tree prior is implicitly defined through a stochastic process, growing trees from their root nodes according to two functions $p_\text{SPLIT}(\eta,T), p_\text{RULE}(\rho\given\eta,T)$, where $\eta$ represents a tree node and $\rho$ a probability distribution over all splitting rules. At each $\eta$, the tree grows (node splits) according to these functions. Various function choices are possible, with \cite{Chipman98} suggesting $p_\text{SPLIT}(\eta,T) = \alpha(1+d_\eta)^{-\beta}$ (hyperparameters $0 < \alpha < 1, \beta \geq 0$ and $d_\eta$ the depth of $\eta$) and $p_\text{RULE}(\rho\given\eta,T) = \frac{1}{J}\frac{1}{n}$ (uniform specification over all predictors $J$ and observations $n$).    
Trees can be conceptualized based on two characteristics: the collection of splitting rules, and the order in which splits occur. The first can be further broken down into the chosen predictors, and their values which partition the space (splitting values), both components of $p_\text{RULE}$. 	Both choice of predictor and splitting value are candidates for sharing of information between groups in a multilevel model, and we will consider each in turn.
    
\paragraph*{Tree prior: Dirichlet-Multinomial model for variable selection}
	We first want to share information about predictors between groups: if one predictor is a good choice for one tree in one group (leading to a better partition of the space compared to another predictor), ideally we should borrow that information when growing a tree for a related group within the hierarchical structure.  When choosing among predictors for related trees, we should be more likely to split on the same predictor that led to a good split in another tree. While we are not explicitly doing variable selection, the same idea is applicable here, first introduced in BART by \cite{Linero18}. We assume there are $K$ total groups with associated trees: $T_1,\dots,T_K,$ and $\bm{T}=\bigcup_{k=1}^KT_k$. In our application, each group is an ecoregion and $K = 9$. Let $\bm{s} = (s_1,\dots,s_J)^T$ be a vector of splitting probabilities shared among all trees: the probability of splitting on predictor $x_j$ is given by $s_j$. Following \cite{Linero18}, we use a Dirichlet prior: $\bm{s} \sim D(\alpha_s/J, \dots, \alpha_s/J),$
	where $\alpha_s$ is a hyperparameter that informs model sparsity. Since the prior is conjugate, the posterior of $\bm{s}$ is also Dirichlet (see Supplementary Material).
	The Gibbs update for $\bm{s}$ includes a count of how many times each predictor was split across all trees. If a predictor has been used more often, its Dirichlet component will have a larger weight, so that predictor will be more likely to be chosen for future splits.
    
\paragraph*{Tree prior: quantile-based splitting values}
    A natural extension to the sharing of information for choice of predictor is to apply a similar model to the splitting values. We could place a Dirichlet prior on all observed values for each predictor, using a multinomial likelihood for the number of times that particular value is chosen, but this would lead to a far bigger tree space to search than is computationally feasible. Instead, we calculate equally spaced quantiles for each predictor's range, using a Dirichlet-Multinomial model on these quantiles to borrow across groups. This is also a useful way of reducing the tree space to search by discretizing continuous predictors (we will limit ourselves to continuous variables).
	Let $Q$ represent the total number of quantiles considered for a variable. Denote the vector of probabilities of splitting at each quantile for predictor $x_j$ by $\bm{q}_j:=(q_{j1},\dots,q_{jQ})$.
	Assume a Dirichlet prior: \(p(\bm{q}_j) \sim D(\alpha_q/Q,\dots,\alpha_q/Q),\) with hyperparameter $\alpha_q$ controlling sparsity among quantile selection. Just as for $\bm{s}$, this produces a Dirichlet posterior (Supplementary Material), with the same interpretation---a quantile which has been split on more often for a given predictor will be more likely to be chosen in future tree splits across the groups.

\paragraph*{Prior and posterior details}
    For simplicity we adopt the conjugate parameter prior for terminal node means and variances (all quantities are allowed to vary by tree $k$, which is not shown for clarity): $\mu_{b}\given\sigma^2,T \sim N(\bar{\mu},\sigma^2/a),\quad
	\sigma^2\given T \sim IG(\nu/2,\nu\lambda/2),$ with hyperparameters $\bar{\mu}, a, \nu, \lambda$. Marginalizing out $\mu_{b}, \sigma^2$ (where $V_b$ is the sample variance in terminal node $b$):
 \begin{align*}
    p(\bm{Y}\given\bm{X},T) &= \frac{ca^{B/2}}{\prod_{b=1}^{B}(n_b+a)^{1/2}}\left (\sum_{b=1}^{B}(v_b+w_b)+\nu\lambda\right )^{-(n+\nu)/2} \\
	v_b = (n_b-1)V_b,\quad w_b &= \frac{n_b a}{n_b+a}(\bar{Y_b}-\bar{\mu})^2,\quad \bar{Y_b} = \frac{1}{n_b}\sum_{i=1}^{n_b}Y_{b,i},\quad b=1,\dots,B.
\end{align*}

  \section{Computational strategy}
\subsection{Metropolis-Hastings algorithm} \label{sec:MH_details}
  The posterior $p(T\given\bm{X},\bm{Y}) \propto p(\bm{Y}\given\bm{X},T)P(T)$ is approximated using a Metropolis-Hastings (MH) algorithm. Each iteration, we propose one of four moves \parencite{Chipman98}:
\begin{enumerate}
    \item Grow: select a terminal node and split into two new terminal nodes.
    \item Prune: select a parent of two terminal child nodes, and remove the child nodes from the tree (so the parent becomes terminal).
    \item Change: assign a different splitting rule to an internal node.
    \item Swap: select internal parent and child nodes, and swap their splitting rules (also swap the other child's rule with the parent if both child nodes have the same rule).
\end{enumerate}
Other moves allow better exploration of the tree space: rotating among nodes \parencite{GramacyLee_rotate,Pratola_perturb}; perturbing splitting rules \parencite{Pratola_perturb}; and growing/pruning more than one node \parencite{KimRockova_twig}. However, we have found the original four sufficient for mixing in combination with parallel tempering (Section \ref{section_PT}). 
We followed \cite{Chipman98} and used uniform proposals; MH ratios are deferred to Supplementary Material. We used the \texttt{partykit} R package \parencite{partykit} for tree implementation.

\subsection{Parallel tempering within Gibbs}
\label{section_PT}
Bayesian CART MCMC algorithms often mix poorly, with chains becoming stuck near local posterior modes and unable to realistically explore the full posterior surface \parencite{Chipman98}. A number of approaches to improve mixing using the concept of tempering have been proposed. The common thread is to run a series of Markov chains with the likelihood ratio raised to various powers, flattening the posterior surface and enabling movement away from local maxima. Information from flatter surfaces is then propagated through to the true posterior, allowing better searching of its surface.
We use a version of parallel tempering (PT) to improve mixing in our multilevel regression trees, which suffer from the same mixing difficulties as other Bayesian CART algorithms. We first review the basic approach to PT.

Let \(1 =\tau_1 \geq \tau_2 \geq \dots \geq \tau_M = 0\) denote a series of temperatures. 
Define a series of (unnormalized) densities by 
	$p_{\tau_m}(T\given\bm{X},\bm{Y})
	\propto p(\bm{Y}\given\bm{X},T)^{\tau_m}P(T).$
MH ratios are adjusted ($g(T^{(\ell)},T^*)=P(T^*\given T^{(\ell)})$ and $T^*, T^{(\ell)}$ denote proposed and current trees, respectively):
	\[ \frac{g(T^*,T^{(\ell)})}{g(T^{(\ell)},T^*)} \frac{p(\bm{Y}\given\bm{X},T^*)^{\tau_m}P(T^*)}{p(\bm{Y}\given\bm{X},T^{(\ell)})^{\tau_m}P(T^{(\ell)})}.\]
 Trees are randomly swapped between adjacent chains with probability $\min\{1,A\}$, where
\begin{align}
		A &=  \frac{p_{\tau_m}(T^{(\ell+1)})p_{\tau_{m+1}}(T^{(\ell)})}{p_{\tau_m}(T^{(\ell)})p_{\tau_{m+1}}(T^{(\ell+1)})} 
		= \frac{p(\bm{Y}\given\bm{X},T^{(\ell+1)})^{\tau_m}p(\bm{Y}\given\bm{X},T^{(\ell)})^{\tau_{m+1}}}{p(\bm{Y}\given\bm{X},T^{(\ell)})^{\tau_m}p(\bm{Y}\given\bm{X},T^{(\ell+1)})^{\tau_{m+1}}}. \label{swap_ratio}
\end{align}
Ideally, lower temperature chains are able to find `better' trees closer to the global maximum of the posterior surface. Such trees are then more likely to move into the true posterior chain having made random swaps between adjacent chains. 

Consider how updating shared hyperparameters $(\bm{s},\bm{q}_{1:J})$ interacts with PT. After updating $K$ related trees, parameters, and hyperparameters at each temperature, we could swap all quantities between adjacent chains using a joint likelihood. In practice, this approach appeared to lead to poor mixing---if hyperparameters are optimized for a particular temperature chain, swapping them between chains is likely prohibitive in achieving better tree space exploration. Instead, we propose a new algorithm to swap between adjacent trees. We only update $(\bm{s},\bm{q}_{1:J}) \sim p((\bm{s},\bm{q}_{1:J})\given\bm{T})$ using the Gibbs updates for $\bm{s}$ (\ref{S-s_update}) and $\bm{q}_j$ (\ref{S-q_update}) once each iteration at the original temperature $\tau_1=1$, corresponding to the true posterior. Conditional on the same current $(\bm{s},\bm{q}_{1:J})$ value for all trees, perform Algorithm \ref{alg:PT_within_Gibbs}.
\begin{algorithm}
\caption{Parallel Tempering within Gibbs}\label{alg:PT_within_Gibbs}
\begin{algorithmic}
\For{$k=1$ \textbf{to} $K$}
    \For{$m=1$ \textbf{to} $M$}
        \State Update $T_k \sim p(T_k\given (\bm{s},\bm{q}_{1:J}),\bm{Y}_k) \propto p(T_k\given (\bm{s},\bm{q}_{1:J}))p(\bm{Y}\given T_k)^{\tau_m}$ at each $\tau_m$ (MH step)
    \EndFor
    \For{$m=1$ \textbf{to} $M-1$}
        \State Randomly swap \textbf{only trees} between chains at $\tau_m$ and $\tau_{m+1}$ with probability $\min(1,A)$, where $A$ follows equation \ref{swap_ratio} but only involving $T_k$ and $\bm{Y}_k$, respectively.
    \EndFor
    \State Retain samples $T_k$ at $\tau_1=1$ only, corresponding to samples from the true posterior.
\EndFor
\end{algorithmic}
\end{algorithm}
By limiting PT swaps to only trees and data from the same group $k$, we are reducing potential contamination from other trees and groups. Since $(\bm{s},\bm{q}_{1:J})$ remains fixed at $\tau_1=1$ even for chains of different temperatures, we are more likely to sample trees from the true posterior at all temperatures.

\section{Simulation study} \label{sim_desc}
In order to test our multilevel regression trees, we ran a simulation study comparing three different models:
\begin{enumerate}
    \item Partial pooling (one tree fit to each group, sharing of information between groups)
    \item No pooling (one tree fit independently to each group, no sharing of information)
    \item Complete pooling (all data pooled, single tree fit, no distinguishing between groups).
\end{enumerate}
We randomly drew four predictors 100,000 times from a $N(0,1)$ distribution, using $Q=10$ per predictor, replicating 100 times with different random seeds. We fit models 1-3 to each data set (300 total simulations). We simulated true trees from the tree prior to avoid the structure of a single true tree driving the results. We set $\alpha=0.95,\beta=0.1$ (ensuring several splits) and $\bm{s}=(0.45,0.4,0.1,0.05)^T$, with $\bm{q}_j$ randomly drawn from the Dirichlet prior ($\alpha_s, \alpha_q=1$). We forced a minimum of 5 observations per terminal node, at least 5 terminal nodes, and a maximum node depth of 8 across the tree. Per simulation, we generated 9 different groups and corresponding trees (the same number of groups as the wildfire data: Section \ref{sec:wildfire}). 

Having simulated tree structures, we next randomly assigned terminal node means at equally spaced points from -20 to 20, according to the number of terminal nodes in the generated tree (example true tree shown in Supplementary Material Figure \ref{S-fig:example_tree}). We used each tree to simulate actual response values, adding noise following a $N(0,0.5^2)$ distribution. We set the number of response values per group, $n_k \sim U(100,2000)$.  Finally, we generated complete training data sets by randomly sampling $n_k$ rows per group, with all remaining observations $(100,000-n_k)$ serving as out-of-sample test data. 
We ran all simulations for 15,000 iterations each, discarding the first 10,000 as burn-in (judged sufficient from tree log-likelihood trace plots). To improve mixing, we employed Algorithm \ref{alg:PT_within_Gibbs}, with five temperature chains spaced sigmoidally. After trial and error, we chose $\nu=10, \lambda = 1, a=10, \alpha_s=\alpha_q=1$. We set $\alpha=0.5, \beta=2$ to give a stronger penalty on deep splits and constrain trees from getting too large; and $\bar{\mu}$ was the empirical mean of each group's training data.

\subsection{Tree posterior point estimators}  \label{medoid_desc}
We used two representative trees from each posterior: Maximum a posteriori (MAP) and medoid trees. Both gave similar results, so we report performance using medoid trees.

\begin{enumerate}
    \item \textbf{MAP trees}. We counted the unique log-likelihood values visited in each posterior, identified the most common value, and chose the first tree with this value. We assume that identical log-likelihoods correspond to the same tree. It is unlikely this assumption will be violated in practice; even a single different split, or a single different observation at a terminal node, can produce a very different log-likelihood value.
\item \textbf{Medoid trees}. Using a tree distance metric \parencite{bertsimas2023}, we calculated the pairwise distances between all trees in the posterior distribution. Weighting each distance by the number of times trees were sampled, the tree that had the minimum total distance from all other trees was designated the medoid tree.
\end{enumerate}

\subsection{Performance metrics}
We chose three metrics to evaluate tree model performance in each simulation. Given the importance of interpretability for our multilevel regression trees, we took into account tree structure when assessing performance, not only prediction. We employed the tree distance metric introduced by \cite{bertsimas2023} that finds the distance between trees by solving a linear programming problem, minimizing the distance between sub structures.

\begin{enumerate}
\item \textbf{Mean Posterior Distance (MPD)}.
All unique log likelihoods were identified, assumed to be in 1:1 correspondence with unique trees. We then calculated the distance of each unique tree from each group's true tree, and finally calculated the mean distance per group.
\item \textbf{Root Mean Squared Error (RMSE)}.
Medoid trees were used to predict out-of-sample test data responses: the mean of each observation's terminal node. Residuals were squared, added, and the square root taken to find the RMSE for the test set.
    \item \textbf{Widely Applicable Information Criterion (WAIC)}.
    Using the posterior samples for $\mu_b$ and $\sigma^2$ (Equations \ref{S-eq:mu_i}, \ref{S-eq:sigma2}), pointwise log-likelihoods were evaluated. WAIC  \parencite{Watanabe2010} was calculated using the \texttt{loo} R package \parencite{loo_package,loo_paper}, according to \cite{loo_paper} ($V_{\ell=1}^L(\cdot)$ is the sample variance operator):
\begin{equation*}
    \widehat{elpd}_{\text{WAIC}} = \sum_{i=1}^n\log \left(\frac{1}{L}\sum_{\ell=1}^Lp(y_i\given\theta^{(\ell)}) \right) -\sum_{i=1}^nV_{\ell=1}^L(\log (\frac{1}{L}\sum_{\ell=1}^L p(y_i\given\theta^{(\ell)})). 
\end{equation*}

\end{enumerate}

\subsection{Results}
Boxplots of each performance metric across the 100 repeated simulations are shown in Figure \ref{fig:simulation_summary} (MPD and RMSE were means across all 9 groups; groupwise boxplots are shown in Supplementary Material, Figs \ref{S-fig:group_MPD} and \ref{S-fig:group_RMSE}). In general, complete pooling performed the worst in every situation---this was unsurprising since true trees were not the same across groups.
\begin{figure}
    \centering
    \includegraphics[width=\linewidth]{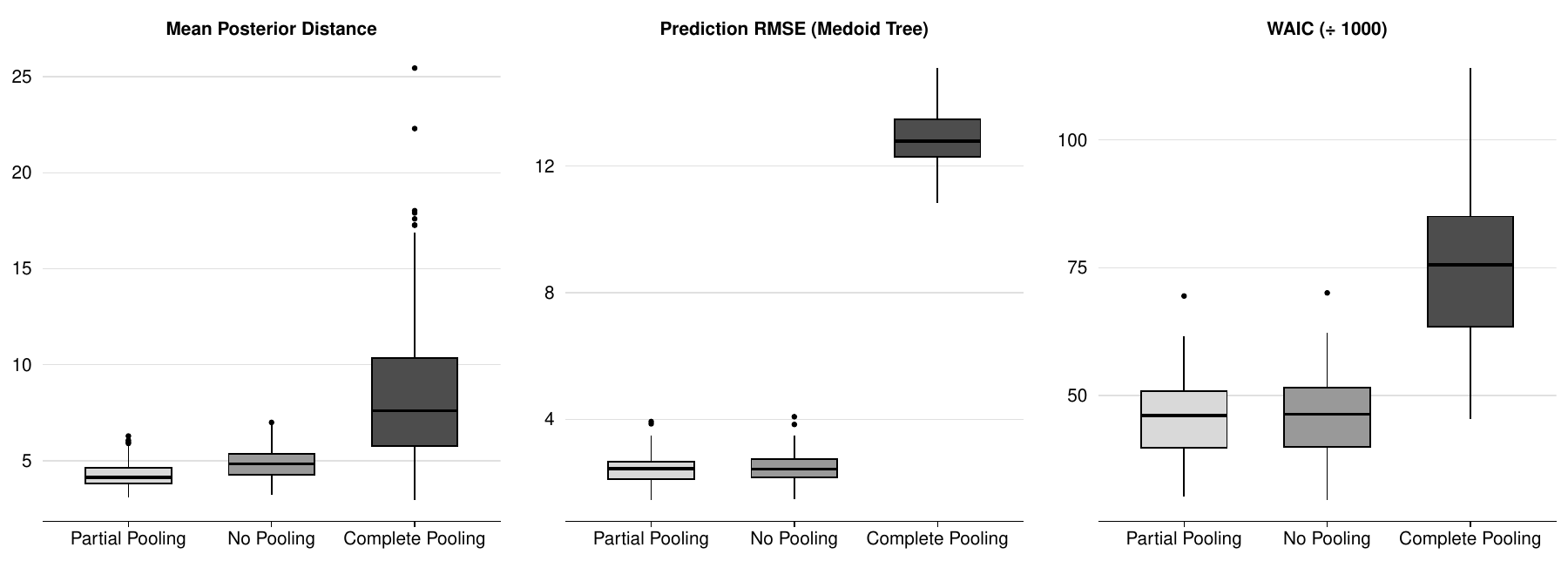}
    \caption{Simulation study performance metrics, replicated 100 times at random seeds. MPD and prediction RMSE values are means across all 9 groups.}
    \label{fig:simulation_summary}
\end{figure}
MPD showed the greatest benefit of sharing of information between groups. Partial pooling often had the lowest MPD, seen in the boxplot being narrower and shifted lower relative to no pooling (Fig. \ref{fig:simulation_summary}, left). These results give us good confidence that sharing variable splitting information results in tree posteriors structurally closer to true trees, compared to no pooling. RMSE (Fig. \ref{fig:simulation_summary}, center) was roughly the same  between no pooling and partial pooling (uniformly worse for complete pooling). While we would like to see partial pooling do better in the prediction task, it is reassuring that it performs no worse than no pooling. WAIC values for partial pooling and no pooling were comparable (Fig. \ref{fig:simulation_summary}, right), with perhaps a slight benefit for partial pooling (the median is slightly shifted down compared to no pooling).

Overall, trees that were partially pooled tended to show the closest structural similarity to the true trees. Metrics based on predictive performance (RMSE, WAIC) did not show as strong an advantage to partial pooling, but comparable performance to no pooling. Given the importance of tree structure and interpretability in our approach, along with the usefulness of posterior summaries only available from partial pooling, we are confident our method can be profitably applied to real data. A variety of other simulations are possible: smaller, larger, or a mixture of sample sizes between groups; and variations on tree generation and associated hyperparameters. However, we believe the simulation setup presented here has sufficiently demonstrated the utility of sharing of information between groups via shared variable splits.

\section{Wildfire data}\label{sec:wildfire}

\subsection{Description of data set}
In order to apply our multilevel regression trees approach to the wildfire problem, we used a subset of the wildfire data assembled for the  2021 Extreme Value Analysis (EVA) conference data challenge \parencite{Opitz2023}. It is reasonable to assume that fires occurring in similar geographic regions would have similar explanatory/contributing factors, providing a useful situation in which to share variable splitting information between groups (ecoregions).

The data consist of wildfires that occurred in the western United States between 1993 and 2015, March to September each year \parencite{Short2017}. Monthly aggregate total burned area (BA) in acres was recorded on a spatial grid of pixels, each approximately 55 km by 55 km. We adjusted BA by dividing by the proportion of that pixel within the US (if the pixel was on a coast or border, less than the full area was able to burn), then added 1 to BA in case of zeros, and log-transformed to have an unconstrained response variable. A summary of all covariates used is shown in Table \ref{tab:covariates}. One set of covariates, land cover, was sourced from the European Union COPERNICUS Climate Data Store. These variables are based on the United Nations Food and Agriculture Organization’s Land Cover Classification System, filtered to 18 different land classification types for the US. Each classification type was given as a proportion, summing to 1 over all types in each pixel. We further aggregated several categories according to common land classification in the American West. The other main covariates were mean monthly climate variables, sourced from ERA5-reanalysis on Land surface \parencite{ERA5}, and altitude mean and standard deviation. All covariates were standardized.
\begin{table}[htbp]
\centering
\caption{Covariates in EVA data \parencite{Opitz2023}, grouped by category.}
\label{tab:covariates}
\small
\begin{tabular}{p{0.28\textwidth} l p{0.5\textwidth}}
\toprule
\textbf{Covariate} & \textbf{Category} & \textbf{Description} \\
\midrule
Forest (needle) & Land cover & tree needleleave evergreen closed to open, tree needleleave evergreen closed \\
Forest (broad) & Land cover & tree broadleaved deciduous closed to open, tree broadleaved deciduous closed, tree mixed \\
Shrub/grass & Land cover & mosaic tree and shrub, shrubland, grassland, sparse vegetation \\
Agriculture & Land cover & cropland rainfed, cropland rainfed herbaceous cover, mosaic cropland, mosaic natural vegetation \\
Non burnable & Land cover & urban, bare areas, water \\
Wetland & Land cover & tree cover flooded fresh or brackish water, shrub or herbaceous cover flooded \\
\midrule
10 m U-component of wind & Climate & the wind speed in Eastern direction (m/s) \\
10 m V-component of wind & Climate & the wind speed in Northern direction (m/s) \\
Dewpoint temperature & Climate & temperature at 2 m from ground to which air must be cooled to become saturated with water vapor, such that condensation ensues (Kelvin) \\
Temperature & Climate & at 2 m from ground (Kelvin) \\
Potential evaporation & Climate & the amount of evaporation of water that would take place if a sufficient source of water were available (m) \\
Surface net solar radiation & Climate & net flux of shortwave radiation; mostly radiation coming from the sun (J/m$^2$) \\
Surface net thermal radiation & Climate & net flux of longwave radiation; mostly radiation emitted by the surface (J/m$^2$) \\
Surface pressure & Climate & (Pa) \\
Evaporation & Climate & of water (m) \\
Precipitation & Climate & (m) \\
Altitude (mean), altitude (SD) & Climate & mean and standard deviation of altitude in each pixel \\
\bottomrule
\end{tabular}
\end{table}
We filtered the data for western states in the US, performing spatial joins to reference data using the R packages \texttt{tigris} \parencite{tigris_paper} and \texttt{sf} \parencite{sfBook,sfPaper}. We then joined to US EPA level II ecoregion shapefile data \parencite{EPA2010NACECEcoLevel2}. Unlike the original data challenge, we did not mask any observations, and instead did a random train/test split of 60-40\% of the data over all years.

\subsection{MCMC settings}
The same three methods as the simulation study were used to fit trees to the fire data: partial pooling (shared variable split information across ecoregions), no pooling (trees fit independently to each ecoregion), and complete pooling (all data pooled across ecoregions and a single tree fit). For each method, MH algorithms were run for 200,000 iterations in parallel. We discarded 190,000 iterations as burn-in, as trace plots of tree log-likelihoods showed a long time to convergence for some ecoregions. PT was used to improve mixing following Section \ref{section_PT}, again with five temperature chains spaced sigmoidally. After some trial and error, we chose $\nu=20, \lambda = 2, a=20, \alpha_s=\alpha_q=1$. Setting $a=20$ helped to grow trees of a reasonable size, while splitting prior hyperparameters were set to $\alpha=0.5, \beta=7$ to give a stronger penalty on deep splits and constrain trees from getting too large. We chose $Q=10$, and $\bar{\mu}$ was set to the sample mean of each ecoregion for each covariate.

\subsection{Results}
Fitted trees varied in complexity across ecoregions, with a general trend that the more data present in an ecoregion, the deeper the trees were with more splits. Complete pooling trees were generally the largest. The primary point estimator we used to summarize tree posteriors was the medoid tree. Since interpretability is a large part of why we use single regression trees rather than ensemble methods, we will study the structure of each partial pooling medoid tree in the next section, before exploring predictive performance. We will then focus on the unique advantage given by the partial pooling approach, analyzing hyperparameters used for sharing of information between ecoregions.

\subsubsection{Tree structure} \label{sec:tree_structure}
\begin{figure}[!ht]
    \centering
    \includegraphics[width=0.8\linewidth, height=\textheight, keepaspectratio]{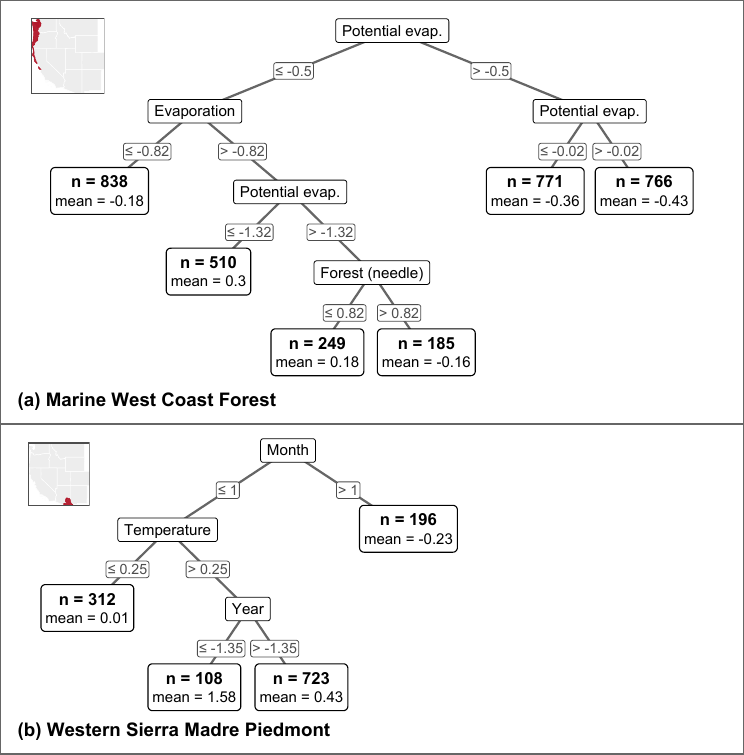}
    \caption{Medoid tree for (a) Marine West Coast Forest and (b) Western Sierra Madre Piedmont (level II ecoregion).}
    \label{fig:medoid_pair1}
\end{figure}

\begin{figure}[!ht]
    \centering
    \includegraphics[width=\linewidth, height=\textheight, keepaspectratio]{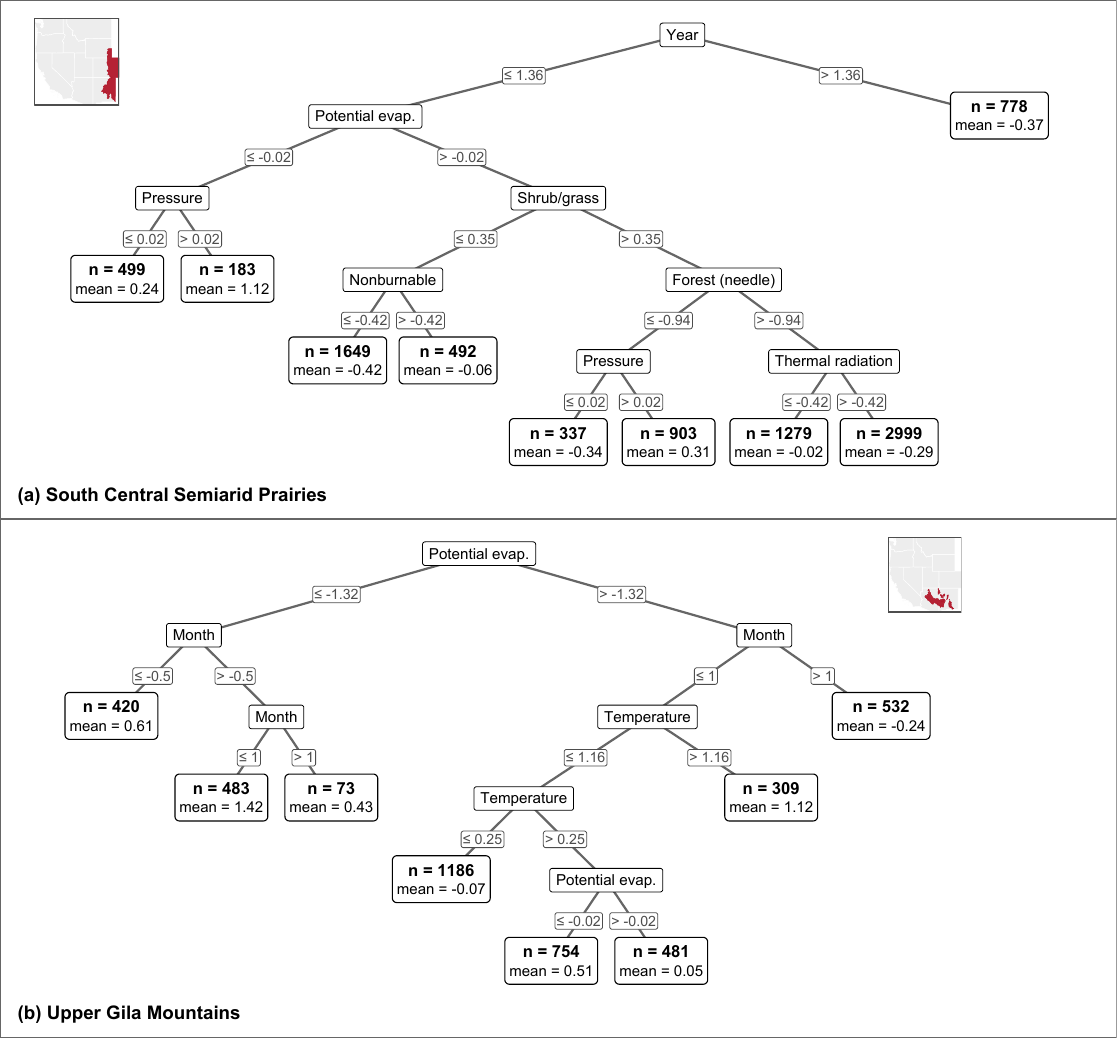}
    \caption{Medoid tree for (a) South Central Semiarid Prairies and (b) Upper Gila Mountains (level II ecoregion).}
    \label{fig:medoid_pair2}
\end{figure}

We calculated medoid trees from each posterior distribution of trees using the same method described in section \ref{medoid_desc}. A selection of partial pooling medoid trees from four ecoregions is shown in Figures \ref{fig:medoid_pair1} - \ref{fig:medoid_pair2}, with the remaining larger ecoregion medoid trees provided as Supplementary Material Figures \ref{S-fig:supp_cold_deserts}-\ref{S-fig:supp_cordillera}.

\paragraph*{Marine West Coast Forest (Fig. \ref{fig:medoid_pair1} (a)).} This tree was among the smallest of the ecoregions, with three variables split on: potential evaporation, evaporation, and forest (needle). This ecoregion covers the coastal regions of primarily Washington, Oregon, and Northern California, known for milder climates year round and precipitation, rather than strong seasonal changes. Potential evaporation split on quantiles 2-4, and evaporation on the 3rd. More negative values indicate more potential evaporation, suggesting that either a large amount, or close to median/little potential evaporation was most predictive of BA. Forest (needle) was the last split, suggesting this land cover type is not as important as climate factors, but that evergreen forests have some importance to wildfires in this ecoregion.
\paragraph*{Western Sierra Madre Piedmont (Fig. \ref{fig:medoid_pair1} (b)).} This ecoregion is very far south, covering a small part of Arizona and New Mexico. Time variables were almost the exclusive splits here, month (August) and year (1995), with temperature ($17^\circ$C) the only climate variable split. It appears that there is strong seasonal behavior in wildfires in these desert areas, perhaps without many climate or land cover factors that predict fire severity. Interestingly the year split here was earlier than most other ecoregion trees with a year split.
\paragraph*{South Central Semiarid Prairies (Fig. \ref{fig:medoid_pair2} (a)).} This ecoregion covers large parts of the midwest, eastern Colorado, Wyoming, and New Mexico. Year (2013) was the first split. Potential evaporation was next on the fourth quantile, indicating a moderate amount of potential evaporation was best at partitioning the space. Pressure was used to split twice, on a value near the median surface pressure. Thermal radiation was a late split. Land cover variables featured heavily, starting with shrub/grass (near 70\%), followed by non burnable and forest (needle), both near 0\%, suggesting either completely non burnable or no non burnable land cover was most important. Given how many different types of land cover occur in this part of the country, it is understandable that land cover variables were important.
\paragraph*{Upper Gila Mountains (Fig. \ref{fig:medoid_pair2} (b)).} This is a smaller ecoregion covering parts of Arizona and New Mexico. No land cover variables were used: all splits were on time and climate variables. Land cover is likely very consistent here, while seasonal patterns are important in predicting BA. Most prominent was potential evaporation, splitting on its 2nd quantile, and again much later on its 4th. Month was used three times, splitting twice on August and once on May. These months' importance aligns with \cite{Westerling_fire_review}: in Arizona and New Mexico, wildfire season tends to begin around May, corresponding to depleted snowpack and lower precipitation, and is over by August, usually earlier than the rest of the West. Temperature was used for two splits, the first of which occurred at around $25^\circ$C (9th quantile): high temperatures were most important in predicting wildfires in this ecoregion, compared to more moderate temperature splits in other ecoregions.

\paragraph*{Mediterranean California.} Month (April) was the most important early split. This ecoregion includes all of greater Los Angeles, known for strong seasonal wildfire patterns, and so we would expect month to be particularly explanatory. Thermal radiation also featured prominently here, being used to split three times. Several land cover types were important: forest (needle), and agriculture initially split on very low values, corresponding to either negligible or almost total land cover proportions of agriculture or evergreen forest being most predictive of fire size. Temperature ($22^\circ$C), non burnable, and year (2013) were late splits. 
\paragraph*{Cold Deserts.} This ecoregion encompasses a vast area of the Western US that is not contiguous, including Washington, Oregon, Idaho, Nevada, Utah, Wyoming, Colorado, New Mexico, and Arizona. The medoid tree was the biggest of all the ecoregions. Potential evaporation was the most prominent split, understandably since there is likely not a sufficient body of water available for most evaporation to occur in desert climates. Temperature ($15^\circ$C) and month (July) were the next highest splits, then all land cover variables, pressure, year, solar radiation, altitude mean and SD. Given how large this area is with more historical fire data, it is likely more patterns were able to be found involving most variables. Even so, being a desert, there is still a stronger preference to split on climate variables like temperature and potential evaporation, variables that could be expected to strongly predict fire size.
\paragraph*{Warm Deserts.} Potential evaporation was the first split here, in an ecoregion covering parts of California, Nevada, Utah, Arizona, and New Mexico. Interestingly, this splitting variable was shared between both Cold and Warm Desert ecoregions, the root of both trees. Year (1995) was an early split. Land cover types were used in many splits, especially agriculture, forest (broad), forest (needle), shrub/grass, and non burnable; like cold deserts, it is likely many different land cover types occur here, and associations with wildfires can be found across all of them. Temperature and both altitude variables were later splits. 
\paragraph*{West-Central Semiarid Prairies.} This ecoregion includes parts of Montana and Wyoming. Land cover variables were very prominent and all were used to split, with forest (needle) being the first. Often, the values split on for land cover were very low, suggesting either almost none or 100\% of that land cover was most predictive of BA. Climate variables were generally later splits, though thermal radiation and temperature were used more than once.
\paragraph*{Western Cordillera.} This is the second biggest ecoregion covering a vast area: parts of California, Oregon, Washington, Idaho, Montana, South Dakota, Wyoming, Utah, Colorado, New Mexico, and generally mountainous territory. Temperature ($17^\circ$C) was the earliest and most important split, with several subsequent splits. Both types of evaporation were split on several times, as well as solar radiation, pressure, and both altitude mean and SD. Land cover variables were less prominent, later splits: forest (needle), and agriculture. Neither month nor year were used, suggesting that seasonality may not play a large role in predicting BA (for example, it is known that wildfires can occur year round in Colorado's Front Range). Alternatively, perhaps stronger seasonal patterns in wildfire behavior at more local levels is contradicted by opposite patterns in a much bigger ecoregion, leading to no clear associations.
\paragraph*{Summary.} We can identify patterns in neighboring ecoregions. Cold Deserts and Western Cordillera are both large areas that often border each other. Similar variables were used to split in both trees: temperature, and potential evaporation (less important in Western Cordillera compared to Cold Deserts, but still prominent). Mediterranean California borders Western Cordillera in many places, and shares some splits (temperature), but seasonality was more predictive in Mediterranean California. South Central Semiarid Prairies borders Western Cordillera in places, and both shared land cover splits, but these were more predictive in the former ecoregion. West Central Semiarid Prairies borders both ecoregions and again here, like South Central Semiarid Prairies, land cover was very prominent. Upper Gila Mountains and Western Cordillera both feature mountainous territory, and both relied more on climate variable splits, with temperature used in both but more important in Western Cordillera. This makes sense given the vast area covered compared to Upper Gila Mountains, where temperature is likely more consistent. Potential evaporation was important in both regions, and evaporation was more important in Western Cordillera.

\subsubsection{Model performance}

Tree distance metrics could not be used to assess model performance on the real data, as there was no true tree to which we could compare each posterior distribution. But based on simulation performance (Fig. \ref{fig:simulation_summary}), we are confident that partial pooling will indeed find trees generally closest to the best tree for each ecoregion, and better than no or complete pooling. To measure predictive performance, we computed three metrics. Prediction error was assessed pointwise using RMSE: calculating each medoid tree's predicted BA in the test data and comparing to true values (Table \ref{tab:RMSE}). We computed 90\% interval coverage by taking the 5\% and 95\% posterior quantiles for predicted BA in the test set, and calculating whether each true BA was within these values (Supplementary Material, Table \ref{S-tab:coverage}). Finally, we calculated WAIC; relative differences using \texttt{loo\_compare} \parencite{loo_compare_paper} are shown in Table \ref{tab:WAIC}. 

For another set of comparisons, we fit classical RPART trees \parencite{rpart} to each ecoregion's training data, cross-validating the complexity parameters. We also fit classical multilevel linear regressions with group-specific intercepts for ecoregions with \texttt{lme4} \parencite{lme4}. Each model was used to predict on the test data; results are also given in Table \ref{tab:RMSE}. Interval estimates for RPART and multilevel linear in Table \ref{S-tab:coverage} are based on bootstrapping and profile likelihoods \parencite{merTools}, respectively.

\begin{table}[htbp]
\centering
\caption{RMSE by ecoregion and method. Lighter indicates the lowest (best) RMSE within each row, darker indicates the highest (worst) RMSE within each row.}
\label{tab:RMSE}
\small
\resizebox{\textwidth}{!}{%
\begin{tabular}{lS[table-format=1.4]S[table-format=1.4]S[table-format=1.4]S[table-format=1.4]S[table-format=1.4]}
\toprule
Ecoregion & {\textbf{Partial Pooling}} & {\textbf{No Pooling}} & {\textbf{Complete Pooling}} & {\textbf{rpart Trees}} & {\textbf{lmer}} \\
\midrule
Cold Deserts & \cellcolor{cbHigh!100!cbLow}{0.8899} & \cellcolor{cbHigh!98!cbLow}{0.8907} & \cellcolor{cbHigh!92!cbLow}{0.8941} & \cellcolor{cbHigh!94!cbLow}{0.8931} & \cellcolor{cbHigh!0!cbLow}{0.9403} \\
Marine West Coast Forest & \cellcolor{cbHigh!100!cbLow}{0.4768} & \cellcolor{cbHigh!78!cbLow}{0.4868} & \cellcolor{cbHigh!0!cbLow}{0.5216} & \cellcolor{cbHigh!79!cbLow}{0.4863} & \cellcolor{cbHigh!10!cbLow}{0.5173} \\
Mediterranean California & \cellcolor{cbHigh!88!cbLow}{1.0316} & \cellcolor{cbHigh!98!cbLow}{1.0202} & \cellcolor{cbHigh!73!cbLow}{1.0471} & \cellcolor{cbHigh!100!cbLow}{1.0185} & \cellcolor{cbHigh!0!cbLow}{1.1266} \\
S. Central Semiarid Prairies & \cellcolor{cbHigh!84!cbLow}{0.8634} & \cellcolor{cbHigh!100!cbLow}{0.8569} & \cellcolor{cbHigh!0!cbLow}{0.8970} & \cellcolor{cbHigh!58!cbLow}{0.8736} & \cellcolor{cbHigh!31!cbLow}{0.8845} \\
Upper Gila Mountains & \cellcolor{cbHigh!100!cbLow}{1.0356} & \cellcolor{cbHigh!51!cbLow}{1.0455} & \cellcolor{cbHigh!0!cbLow}{1.0561} & \cellcolor{cbHigh!55!cbLow}{1.0448} & \cellcolor{cbHigh!75!cbLow}{1.0407} \\
Warm Deserts & \cellcolor{cbHigh!100!cbLow}{0.8011} & \cellcolor{cbHigh!89!cbLow}{0.8085} & \cellcolor{cbHigh!80!cbLow}{0.8144} & \cellcolor{cbHigh!85!cbLow}{0.8114} & \cellcolor{cbHigh!0!cbLow}{0.8691} \\
W-Central Semiarid Prairies & \cellcolor{cbHigh!100!cbLow}{0.7667} & \cellcolor{cbHigh!94!cbLow}{0.7687} & \cellcolor{cbHigh!22!cbLow}{0.7930} & \cellcolor{cbHigh!69!cbLow}{0.7772} & \cellcolor{cbHigh!0!cbLow}{0.8006} \\
Western Cordillera & \cellcolor{cbHigh!100!cbLow}{0.8302} & \cellcolor{cbHigh!92!cbLow}{0.8321} & \cellcolor{cbHigh!92!cbLow}{0.8322} & \cellcolor{cbHigh!76!cbLow}{0.8359} & \cellcolor{cbHigh!0!cbLow}{0.8536} \\
W. Sierra Madre Piedmont & \cellcolor{cbHigh!63!cbLow}{1.1769} & \cellcolor{cbHigh!100!cbLow}{1.1496} & \cellcolor{cbHigh!0!cbLow}{1.2239} & \cellcolor{cbHigh!27!cbLow}{1.2036} & \cellcolor{cbHigh!23!cbLow}{1.2067} \\
\bottomrule
\end{tabular}%
}
\end{table}

\begin{table}[htbp]
\centering
\caption{Comparison of Widely Applicable Information Criterion (WAIC) by pooling method, relative to the best performing model set to 0 elpd (expected log pointwise predictive density). $\Delta\mathrm{elpd}$ is worse when lower (more negative).}
\label{tab:WAIC}
\small
\begin{tabular}{lS[table-format=-4.1]S[table-format=3.1]}
\toprule
Model & {$\Delta$\textbf{elpd}} & {\textbf{s.e. (difference)}} \\
\midrule
Partial Pooling & 0.0 & 0.0 \\
No Pooling & -1677.1 & 133.9 \\
Complete Pooling & -2674.5 & 195.3 \\
\bottomrule
\end{tabular}
\end{table}

Multilevel regression trees consistently performed among the best of all methods based on RMSE, with the smallest RMSE values often occurring for partial pooling (Table \ref{tab:RMSE}). No pooling trees also performed very well and often better based on RMSE, as did some of the classical RPART trees. Complete pooling trees usually did poorly on RMSE, as did the linear model fit to each ecoregion with sharing of information. Interval coverage was often best (closest to 90\%) for partial pooling, comparable with no pooling, and generally worse (further from 90\%) for complete pooling, rpart and multilevel linear (Table \ref{S-tab:coverage}).
Comparing the tree methods using WAIC showed partial pooling as the best performing by a good margin (Table \ref{tab:WAIC}). No pooling and complete pooling had expected log pointwise predictive densities several standard errors below that of partial pooling, giving good evidence that better performance was attributable to more than just noise. While the \texttt{waic()} function produced warning messages for several points, in each case these were a small proportion of the data.  Like the simulation study, WAIC values are roughly comparable for no pooling and partial pooling, and both demonstrably better than complete pooling. 

\subsubsection{Shared hyperparameters from partial pooling} \label{subsec:s_q_interpretation}
\begin{figure}[!ht]
    \centering
    \includegraphics[width=0.8\linewidth]{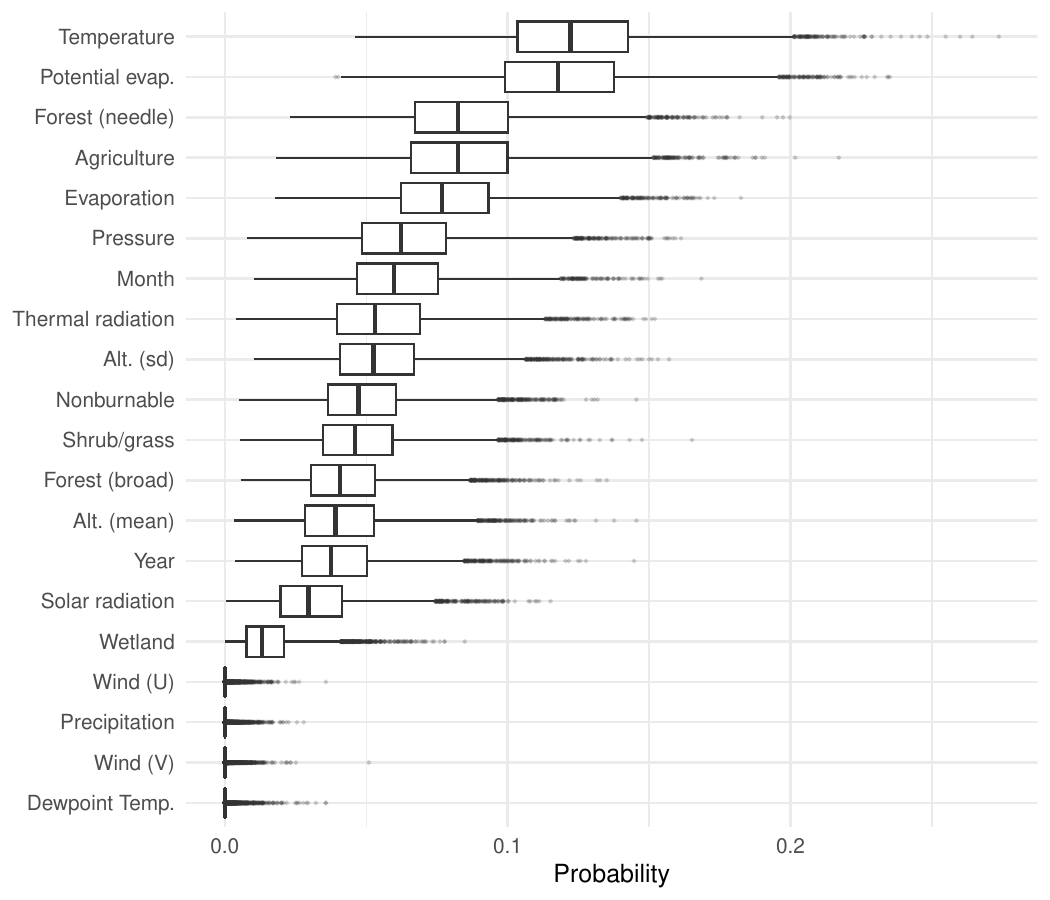}
    \caption{Boxplot showing posterior probabilities of splitting on each variable (vector $\bm{s}$).}
    \label{fig:s_boxplot}
\end{figure}

\begin{figure}
    \centering
    \includegraphics[width=0.8\linewidth]{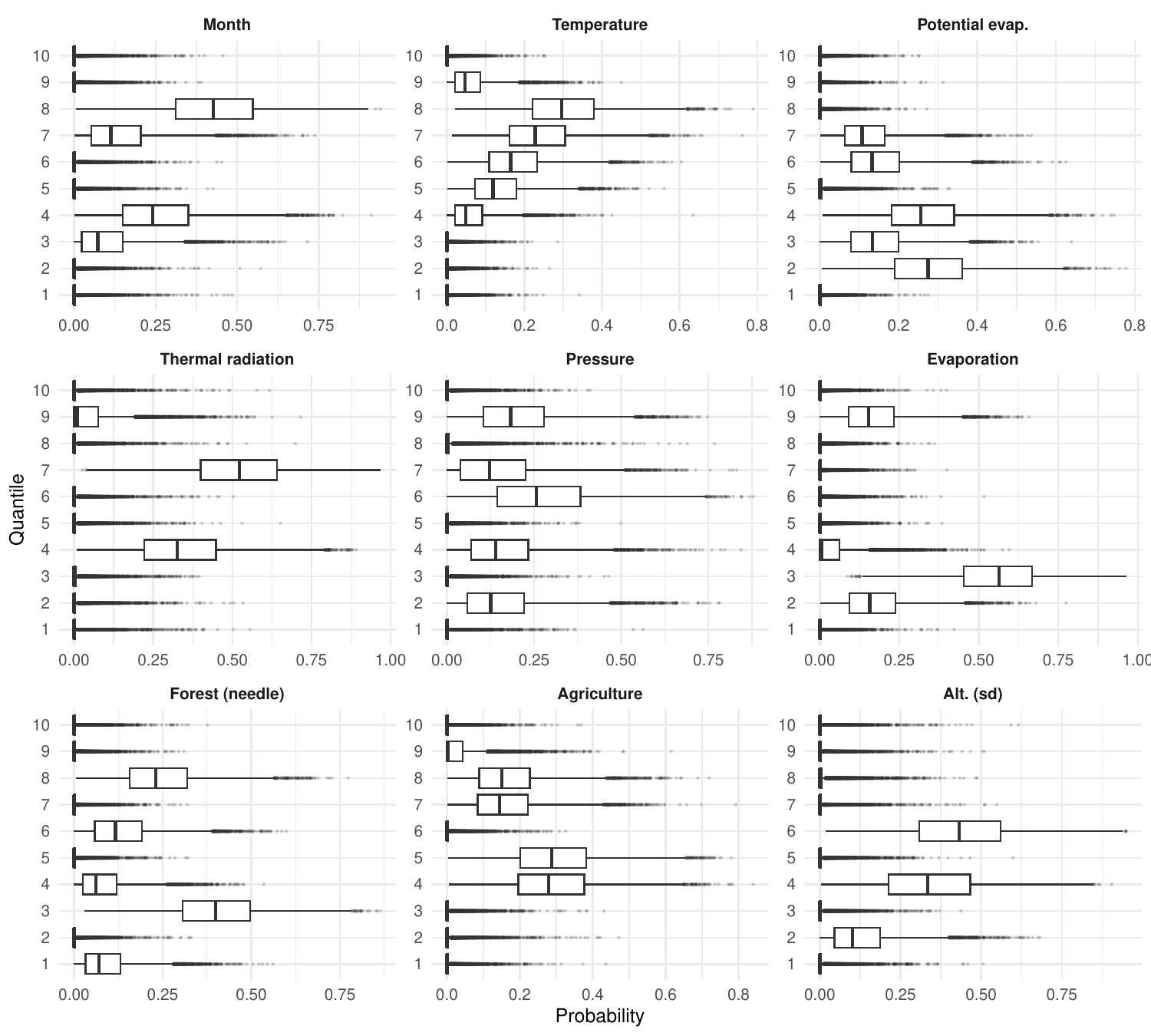}
    \caption{Boxplot showing posterior probabilities of splitting variable quantiles, for the top nine variables most likely to split (vector $\bm{q}_j$).}
    \label{fig:q_boxplot}
\end{figure}

The partial pooling method provides us with very useful outputs with no equivalent from the other models: the posterior distributions for both $\bm{s}$ and $\bm{q}_j$ vectors. These give some indication of variable importance: a higher value for a component of each respective vector reflects a higher probability of splitting on that variable or quantile when growing trees. The posterior probabilities for $\bm{s}$ are shown in Figure \ref{fig:s_boxplot}, and the posterior probabilities for each of the nine most popular $j$ variables $\bm{q}_j$ are shown in Figure \ref{fig:q_boxplot}.

Across all ecoregions, the two most common variables to split on were temperature and potential evaporation (Fig. \ref{fig:s_boxplot}). Clearly temperature (2m from the ground) is highly predictive of the area burned in wildfires. Potential evaporation, defined as the amount of evaporation of water that would take place if a sufficient source of water were available, is more predictive than actual evaporation, a less frequently chosen splitting variable. Given how much of the western ecoregions are essentially dry, desert climates (warm and cold, and mountains), it is reasonable that potential evaporation would be more important, as sufficient sources of water may not be readily available. 

Examining these variables further, the most likely splitting value for temperature was the 8th quantile, corresponding to around $22^\circ$C, with decreasing probabilities as quantiles both increase and decrease (Fig. \ref{fig:q_boxplot}). Higher temperatures were most predictive of fire size, though the highest and lowest temperatures were not very predictive. Less of a pattern was visible in the potential evaporation splitting values, with the second and fourth quantiles most likely to split, but odd quantiles less likely. Actual evaporation was most likely to split on the third quantile. For both evaporation measures, more negative values indicate greater evaporation, and so evaporation nearer its extremes (actual or potential) was more predictive of wildfires.

Two variables that were also quite likely to be chosen as splits were land cover types: forest (needle) and agriculture (Fig. \ref{fig:s_boxplot}).  The former covers evergreen forests of a range of canopies found across western ecoregions, presumably a much greater fire risk with burnable material year-round than deciduous trees where leaves are lost periodically. Indeed, the forest (broad) classification includes deciduous trees and was one of the least likely variables to split for predicting BA. Agriculture land is also found extensively throughout these ecoregions, and perhaps does not provide sufficient fuel for burning. Wetland, non burnable, and shrub/grass land cover types were much less likely to split. 

Both forest (needle) and agriculture showed some pattern in the values used for splitting (Fig. \ref{fig:q_boxplot}). For agriculture, the fourth and fifth quantiles were the most likely splitting values, with seventh through ninth also used, but the lowest and highest quantiles had much lower posterior probabilities. This suggests that across western ecoregions, having a mixture of land cover types with up to 30\% agricultural land cover was most predictive of wildfire BA. By contrast, having almost exclusive (10th quantile) or little (1st-3rd quantile)  agricultural land cover was much less so. Forest (needle) showed a different pattern, with the 3rd, 6th and 8th quantiles often chosen. It appears that having lower, median, and higher proportions of evergreen forest of a variety of canopy cover was most predictive of fire, but not 100\%.

Month was also highly likely to be chosen for splitting, with year less so (Fig. \ref{fig:s_boxplot}). The most predictive splitting values (Fig. \ref{fig:q_boxplot}) for month were the 4th (May), 7th (July), and 8th (August) quantiles. April also somewhat likely to be chosen, and March, June, and September almost not at all. Across all western ecoregions, being in the peak summer months is clearly an important predictive factor in wildfire BA. Finally, surface pressure, thermal radiation, and altitude standard deviation were somewhat likely to be used for splitting (Figs. \ref{fig:s_boxplot}, \ref{fig:q_boxplot}). Surface pressure was most likely to split on the 6th quantile. Thermal radiation was most likely to split on the 4th and 7th quantiles. Altitude SD was most likely to split on the 6th quantile, followed by the 4th; interestingly the 5th quantile was very unlikely to be used. Mean altitude was less predictive of fire size compared to standard deviation. Precipitation, dewpoint temperature, and measures of wind speed were very unlikely to be used for splitting in any ecoregion.

\section{Discussion}

In this work, we have applied a multilevel regression tree approach to the analysis of area burned in wildfires in the western United States. This model enables sharing of variable splitting information among ecoregions within a broadly similar overall area of the US. In addition to providing comparable or even better predictive performance of wildfire size relative to no pooling models, we obtain the distinct advantages of increased model interpretability and understanding of factors associated with wildfires in related ecoregions. In particular, we know that temperature, potential evaporation, and forest (needle) land cover are the most important factors in predicting wildfires across the western US ecoregions. We can also relate variables and values split on to characteristics of each ecoregion, such as those where climate is more seasonal. Such understanding, while intuitive and perhaps could be hypothesized, can only be achieved through a multilevel regression model. The use of a regression tree adds the benefits of ease of interpretation and incorporation of complex interactions that would be harder to capture in other models.

Additionally, our simulation study suggests that partial pooling is more likely to find the `best' tree structurally in each ecoregion (compared to no pooling) in terms of MPD from an assumed true tree. This structural advantage indicates that sharing of information between trees extends benefits beyond enhanced understanding of predictors. The best regression trees in similar ecoregions are indeed structurally similar, and we can exploit this fact both for prediction and understanding of wildfires. Rather than being mutually exclusive, better predictive performance goes hand in hand with these structural gains, as illustrated through WAIC comparisons on the wildfire data. Partial pooling achieved the best WAIC values relative to no pooling and complete pooling.

We also applied a novel computational approach to improve MCMC mixing within multilevel models with shared hyperparameters: PT within Gibbs. By conditioning on the current values of shared hyperparameters at the true posterior temperature chain, we can more efficiently explore the true posterior through random swaps with chains at other temperatures. Given the challenges of efficiently exploring the posterior space when fitting trees extends as far back as \cite{Chipman98}, our contribution provides a low-cost, easy to implement way to achieve better mixing in Bayesian tree inference.

In future, we could extend the multilevel tree model to include categorical response variables (classification trees), with only minor modifications necessary. We could also explicitly look at wildfire forecasting rather than doing a train/test random split across all years. The insights obtained through our sharing of information approach could be applied to wildfire mitigation efforts, through prioritizing contributing factors to wildfire severity for targeting. Lastly, we could explore a stronger approach to partial pooling between trees, beyond just sharing variable splitting information---sharing actual tree structures/substructures between groups.

\spacingset{1}

\section{Disclosure statement}\label{disclosure-statement}

The authors have no conflicts of interest to declare.

\section{Data Availability Statement}\label{data-availability-statement}

Deidentified data and code will be made available.


\clearpage
\setcounter{equation}{0}
\setcounter{figure}{0}
\setcounter{table}{0}
\setcounter{section}{0}
\renewcommand{\theequation}{S\arabic{equation}}
\renewcommand{\thefigure}{S\arabic{figure}}
\renewcommand{\thetable}{S\arabic{table}}
\renewcommand{\thesection}{S\arabic{section}}

\def\spacingset#1{\renewcommand{\baselinestretch}%
	{#1}\small\normalsize} \spacingset{1}
\phantomsection\label{supplementary-material}
\bigskip

\begin{center}

	{\large\bf SUPPLEMENTARY MATERIAL}
	
\end{center}

\spacingset{1.8} 

\section{Posterior derivations}
\subsection{Gibbs updates for terminal node means and variances}
In order to evaluate pointwise log-likelihoods (necessary for calculating some information criteria), we derive Gibbs updates for each $\mu_b$ and $\sigma^2$.
{\scriptsize\selectfont
	\begin{align*}
		p(\mu_{1},\dots,\mu_{B},\sigma^2\given \bm{Y},\bm{X},T) &\propto \prod_{b=1}^{B}\prod_{i=1}^{n_b}p(y_{b,i}\given \mu_b,\sigma^2)p(\mu_b\given \sigma^2,T)p(\sigma^2\given T)
		\\ &\propto\prod_{b=1}^{B}\prod_{i=1}^{n_b} \sigma^{-1}\exp\left(-\frac{(y_{b,i}-\mu_{b})^2}{2\sigma^2}\right)\cdot\sigma^{-1}\sqrt{a}\exp\left(-\frac{a(\mu_{b}-\bar{\mu})^2}{2\sigma^2}\right)\cdot\sigma^{-2(\nu/2+1)}\exp\left(-\frac{\nu\lambda}{2\sigma^2}\right)
		\\ &\propto\prod_{b=1}^{B} \sigma^{-n_b}\exp\left(-\frac{1}{2\sigma^2}\sum_{i=1}^{n_b}(y_{b,i}-\mu_{b})^2\right)\cdot\sigma^{-1}\sqrt{a}\exp\left(-\frac{a(\mu_{b}-\bar{\mu})^2}{2\sigma^2}\right)\cdot\sigma^{-2(\nu/2+1)}\exp\left(-\frac{\nu\lambda}{2\sigma^2}\right)
		\\ \implies p(\mu_b\given \mu_{-b},\sigma^2,\bm{Y},\bm{X},T) &\propto \exp\left(\frac{1}{2\sigma^2}\sum_{i=1}^{n_b}(2y_{b,i}\mu_{b}-\mu_{b}^2)\right)\exp\left(-\frac{a(\mu_{b}^2-2\bar{\mu}\mu_{b})}{2\sigma^2}\right)
		\\ &\propto \exp\left\{-\frac{1}{2}\left[\frac{\mu_{b}^2}{\sigma^2}\left(n_b+a\right)-2\mu_{b}\left(\frac{1}{\sigma^2}\sum_{i=1}^{n_b}y_{b,i}+\frac{a}{\sigma^2}\bar{\mu}\right)\right]\right\}
		\\ &\propto \exp\left\{-\frac{1}{2}\left[\mu_{b}^2\left(\frac{n_b}{\sigma^2}+\frac{a}{\sigma^2}\right)-2\mu_{b}\left(\sum_{i=1}^{n_b}\frac{y_{b,i}}{\sigma^2}+\frac{a}{\sigma^2}\bar{\mu}\right)\right]\right\}
		\\ p(\sigma^2\given \bm{\mu},\bm{Y},\bm{X},T) &\propto  \sigma^{-\sum_{b=1}^{B}n_b}\exp\left(-\frac{1}{2\sigma^2}\sum_{b=1}^{B}\sum_{i=1}^{n_b}(y_{b,i}-\mu_{b})^2\right)\cdot\sigma^{-B}\exp\left(-\frac{a}{2\sigma^2}\sum_{b=1}^{B}(\mu_{b}-\bar{\mu})^2\right)\cdot\sigma^{-2(\nu/2+1)}\exp\left(-\frac{\nu\lambda}{2\sigma^2}\right)
		\\ &\propto \sigma^{-2\left(\frac12\sum_{b=1}^{B}n_b + \frac12B + \frac\nu2+1\right)}\exp\left\{-\frac{1}{\sigma^2}\left[\frac{1}{2}\left(\sum_{b=1}^{B}\sum_{i=1}^{n_b}(y_{b,i}-\mu_{b})^2+a\sum_{b=1}^{B}(\mu_{b}-\bar{\mu})^2 +\nu\lambda\right)\right]\right\}.
	\end{align*}
}

\begin{align}
	\implies p(\mu_b\given \mu_{-b},\sigma^2,\bm{Y},\bm{X},T)  &\sim N\left ((n_b\bar{Y_b}/\sigma^2+a/\sigma^2\bar{\mu})(n_b/\sigma^2+a/\sigma^2)^{-1},(n_b/\sigma^2+a/\sigma^2)^{-1}\right ) \label{S-eq:mu_i}\\
	p(\sigma^2\given \mu_1,\dots,\mu_B,\bm{Y},\bm{X},T)  &\sim IG\left (1/2\sum_{b=1}^{B}n_b + 1/2B + \nu/2, 1/2(\sum_{b=1}^{B}\sum_{i=1}^{n_b}(y_{i,b}-\mu_{b})^2+a\sum_{b=1}^{B}(\mu_{b}-\bar{\mu})^2 +\nu\lambda)\right ). \label{S-eq:sigma2}
\end{align}

\subsection{Variable splitting probabilities} \label{S-sec:s_derivation}
Within a given tree $k$, let $t_{jc}$ denote the $c^{th}$ splitting value for a split on variable $x_j$, i.e. the partition induced by $\{x_j\leq t_{jc}\},\{x_j > t_{jc}\}.$ Predictor $x_j$ has $d_{jk}$ total splits. Let $\bm{t}_{j_k} = (t_{j1},\dots,t_{jd_{jk}})^T,\bm{t}_k=(\bm{t}_{1_k},\dots,\bm{t}_{J_k})^T,$ and $\bm{t}=(\bm{t}_1,\dots,\bm{t}_{k})^T$. 

Following \cite{Linero18}, the conditional posterior of $\bm{s}$ can be derived as follows:
\begin{align*}
	p(\bm{s}\given \bm{T},\bm{X},\bm{Y}) &= \frac{p(\bm{X},\bm{Y}\given \bm{T},\bm{s})P(\bm{T}\given \bm{s})p(\bm{s})}{p(\bm{T},\bm{X},\bm{Y})}
	\\ &=\frac{p(\bm{X},\bm{Y}\given \bm{T})P(\bm{T}\given \bm{s})p(\bm{s})}{p(\bm{T},\bm{X},\bm{Y})}
	\\ &\propto P(\bm{T}\given \bm{s})p(\bm{s}),
\end{align*}
since the likelihood $p(\bm{X},\bm{Y}\given \bm{T},\bm{s}) \propto P(\bm{Y}\given \bm{X},\bm{T})$ does not actually depend on $\bm{s}$.
\\ In the tree prior $P(T_k)$, the only component involving splitting probabilities is $p_\text{RULE}(\rho\given \eta,T_k)$:  \begin{align*}
	p_\text{RULE}(\rho\given \eta,T_k) &= P(c^{th}\text{ value},j^{th}\text{ variable}\given \eta,T_k) 
	\\ &=P(c^{th}\text{ value}\given j^{th}\text{ variable},\eta,T_k)P( j^{th}\text{ variable}\given \eta,T_k)
	\\&\propto s_j.
\end{align*}
Over all trees and letting $\mathcal{S}_k$ denote the set of internal nodes in tree $k$, we can write 
\begin{align*}
	P(\bm{T}\given \bm{s}) &\propto \prod_{k=1}^K\prod_{\eta \in \mathcal{S}_k}p_{\text{RULE}}(\rho\given \eta,T_k) 
	\\ &\propto \prod_{k=1}^K\prod_{j=1}^{J}s_j^{d_{jk}},
\end{align*}
which is a multinomial likelihood. 
\begin{equation}
	p(\bm{s}\given \bm{T},\bm{X},\bm{Y}) \propto \prod_{k=1}^K\prod_{j=1}^{J}s_j^{d_{jk}+\alpha_s/J} \sim D\left(\alpha_s/J+\sum_{k=1}^K d_{1k}, \dots, \alpha_s/J+\sum_{k=1}^K d_{Jk}\right). \label{S-s_update}
\end{equation}

\subsection{Quantile-based splitting values} \label{S-sec:q_derivation}

Just as for $\bm{s},$ in the tree prior $P(T_k)$, the only component involving variable splitting probabilities is $p_\text{RULE}(\rho\given \eta,T)$:  
\begin{align*}
	p_\text{RULE}(\rho\given \eta,T_k) &= P(\text{variable }j\text{ split}\given \eta,T_k)P(c ^{th}\text{ quantile, split variable }x_j) 
	\propto q_{jc}.
\end{align*}
Just as for the splitting probabilities, the vector $\bm{q}_j$ does not actually affect the tree likelihood. Let $r_{jc}$ denote the number of times $q_{jc}$ is used to split in all trees.
The posterior simplifies to
\begin{align*}
	p(\bm{q}_j\given \bm{T},\bm{X},\bm{Y}) &\propto P(\bm{T}\given \bm{q}_j)p(\bm{q}_j)
	\\ &\propto \prod_{c=1}^{Q}q_{jc}^{r_{jc}}q_{jc}^{\alpha_q/Q},
\end{align*}
and so \begin{align}
	\bm{q}_j\given \bm{T},\bm{X},\bm{Y} \sim D(\alpha_q/Q+r_{j1},\dots,\alpha_q/Q+r_{jQ}). \label{S-q_update}
\end{align}

\subsection{Normal-normal splitting values}
For a continuous predictor $x_j$, another natural approach to split value selection is to place a continuous prior over the predictor's range. For this model only, we assume there exist group-level means $\mu_{jk}$ and overall predictor means $\mu_j$, updated in each iteration. We assume a normal prior for splitting values, $p(t_{jc}\given \mu_{jk},\mu_{j},\sigma_i^2) \sim N(\mu_{jk},\sigma^2_i)$. Unlike for $\bm{s}$, there is no simple conjugate form for the posterior for $\bm{t}$. Performance with this model was generally poor, and so we did not apply it to the wildfire data.

\begin{align*}
	t_{jc}\given \mu_{jk},\sigma_{i1}^2 &\sim N(\mu_{jk}, \sigma^2_{i1})
	\\ \mu_{jk}\given \mu_j,\sigma^2_{i2} &\sim N(\mu_j, \sigma^2_{i2}) 
	\\ \mu_j &\sim N(\mu_0,\sigma^2_0)
	\\ p(\sigma^2_{i1}) &\propto IG(a_1,b_1)
	\\ p(\sigma^2_{i2}) &\propto IG(a_2,b_2).
\end{align*}

{\scriptsize\selectfont
	\begin{align*} p(\bm{s},\bm{t},T,\bm{X},\bm{Y},\mu_{j},\mu_{jk},\sigma_{i1}^2,\sigma_{i2}^2) &\propto p(T\given \bm{s},\bm{t},\bm{X},\bm{Y},\mu_{j},\mu_{jk},\sigma_{i1}^2,\sigma_{i2}^2)p(\bm{t}\given \mu_{j},\mu_{jk},\sigma_{i1}^2,\sigma_{i2}^2)p(\mu_{jk}\given \mu_j,\sigma_{i2}^2)p(\mu_{j})p(\sigma_{i1}^2)p(\sigma_{i2}^2) \nonumber
		\\&\propto p(\bm{t}\given \mu_{jk},\sigma^2_{i1})p(\mu_{jk}\given \mu_j,\sigma^2_{i2})p(\mu_{j})p(\sigma_{i1}^2)p(\sigma_{i2}^2) \nonumber
		\\ &\propto \prod_{k=1}^{K}\prod_{i=1}^{p}\prod_{j=1}^{d_{ik}}[p(t_{jc}\given \mu_{jk},\sigma_{i1}^2)]p(\mu_{jk}\given \mu_j,\sigma^2_{i2})p(\mu_{j})p(\sigma_{i1}^2)p(\sigma_{i2}^2) \nonumber
		\\ &\propto \prod_{k=1}^{K}\prod_{i=1}^{p}\left[\prod_{j=1}^{d_{ik}}\sigma^{-1}_{i1}\exp\left(-\frac{(t_{jc}-\mu_{jk})^2}{2\sigma_{i1}^2}\right)\right]\sigma^{-1}_{i2}\exp\left(-\frac{(\mu_{jk}-\mu_{j})^2}{2\sigma_{i2}^2}\right)\cdot \nonumber \\ &\qquad\qquad\qquad \sigma^{-1}_0\exp\left(-\frac{(\mu_{j}-\mu_{0})^2}{2\sigma_0^2}\right)\sigma_{i1}^{-2(a_1+1)}\exp\left(-\frac{b_1}{\sigma_{i1}^2}\right)\sigma_{i2}^{-2(a_2+1)}\exp\left(-\frac{b_2}{\sigma_{i2}^2}\right) \nonumber
		\\ &\propto \prod_{k=1}^{K}\prod_{i=1}^{p}\exp\left\{-\frac{1}{2}\left[\frac{1}{\sigma_{i1}^2}\left(\sum_{j=1}^{d_{ik}}(t_{jc}-\mu_{jk})^2-2b_1\right)+\frac{1}{\sigma_{i2}^2}\left((\mu_{jk}-\mu_{j})^2-2b_2\right)+\frac{1}{\sigma_0^2}(\mu_{j}-\mu_{0})^2\right]\right\} \nonumber
		\\ \implies p(\mu_{jk}\given \bm{t},\mu_j,\sigma_i^2) &\propto \exp\left\{-\frac{1}{2}\left[\frac{1}{\sigma_{i1}^2}\sum_{j=1}^{d_{ik}}(\mu_{jk}^2-2\mu_{jk}t_{jc})+\frac{1}{\sigma_{i2}^2}(\mu_{jk}^2-2\mu_{jk}\mu_{j})\right]\right\} \nonumber
		\\ &\propto \exp\left\{-\frac{1}{2}\left[\left(\frac{d_{ik}}{\sigma_{i1}^2}+\frac{1}{\sigma_{i2}^2}\right)\mu_{jk}^2-2\mu_{jk}\left(\frac{\mu_{j}}{\sigma_{i2}^2}+\frac{1}{\sigma_{i1}^2}\sum_{j=1}^{d_{ik}}t_{jc}\right)\right]\right\} \nonumber
		\\ \implies \mu_{jk}\given \bm{t},\mu_j,\sigma_{i1}^2,\sigma_{i2}^2 &\sim N\left(\frac{\frac{\mu_{j}}{\sigma_{i2}^2}+\frac{1}{\sigma_{i1}^2}\sum_{j=1}^{d_{ik}}t_{jc}}{\frac{d_{ik}}{\sigma_{i1}^2}+\frac{1}{\sigma_{i2}^2}}, \frac{1}{\frac{d_{ik}}{\sigma_{i1}^2}+\frac{1}{\sigma_{i2}^2}}\right).
	\end{align*}
}
Similarly, to update $\mu_{j}$:
\begin{align*}
	p(\mu_{j}\given \mu_{jk},\sigma_{i2}^2,\sigma_0^2) &\propto \exp\left\{-\frac{1}{2}\left[\frac{1}{\sigma_{i2}^2}\sum_{k=1}^{K}(\mu_{j}^2-2\mu_{jk}\mu_{j})+\frac{1}{\sigma_0^2}(\mu_{j}^2-2\mu_{j}\mu_{0})\right]\right\} \nonumber
	\\ &\propto \exp\left\{-\frac{1}{2}\left[\left(\frac{K}{\sigma_{i2}^2}+\frac{1}{\sigma_0^2}\right)\mu_{j}^2-2\mu_{j}\left(\frac{\mu_{0}}{\sigma_{0}^2}+\frac{1}{\sigma_{i2}^2}\sum_{k=1}^{K}\mu_{jk}\right)\right]\right\} \nonumber
	\\ \implies \mu_{j}\given \mu_{jk},\sigma_{i2}^2,\sigma_0^2 &\sim N\left (\frac{\frac{\mu_{0}}{\sigma_0^2}+\frac{1}{\sigma_{i2}^2}\sum_{k=1}^{K}\mu_{jk}}{\frac{K}{\sigma_{i2}^2}+\frac{1}{\sigma_{0}^2}}, \frac{1}{\frac{K}{\sigma_{i2}^2}+\frac{1}{\sigma_{0}^2}}\right ).
\end{align*}
Gibbs updates for $\sigma^2_{i1}, \sigma_{i2}^2:$
\begin{align*}
	p(\sigma_{i1}^2\given \bm{t},\mu_{j},\mu_{jk})&\propto
	\sigma_{i1}^{-2a_1-2}\exp\left(-\frac{b_1}{\sigma_{i1}^2}\right)\prod_{k=1}^{K}\sigma_{i1}^{-d_{ik}}\exp\left\{-\frac{1}{2}\left[\frac{1}{\sigma_{i1}^2}\left(\sum_{j=1}^{d_{ik}}(t_{jc}-\mu_{jk})^2\right)\right]\right\} \nonumber
	\\ &\propto (\sigma_{i1}^{2})^{-(\sum_{k=1}^{K}d_{ik}/2+a_1)-1}\exp\left(-\frac{1}{\sigma_{i1}^2}\left[\frac{1}{2}\sum_{k=1}^{K}\sum_{j=1}^{d_{ik}}(t_{jc}-\mu_{jk})^2+b_1\right]\right) \nonumber
	\\ \implies \sigma_{i1}^2\given \bm{t},\mu_{j},\mu_{jk} &\sim IG\left (\frac{1}{2}\sum_{k=1}^{K}d_{ik}+a_1, \frac{1}{2} \sum_{k=1}^{K}\sum_{j=1}^{d_{ik}}(t_{jc}-\mu_{jk})^2+b_1 \right ) \\
	p(\sigma_{i2}^2\given \bm{t},\mu_{j},\mu_{jk})&\propto
	\sigma_{i2}^{-2a_2-2}\exp\left(-\frac{b_2}{\sigma_{i2}^2}\right)\prod_{k=1}^{K}\sigma_{i2}^{-1}\exp\left\{-\frac{1}{2}\left[\frac{1}{\sigma_{i2}^2}(\mu_{jk}-\mu_{j})^2\right]\right\} \nonumber
	\\ &\propto (\sigma_{i2}^{2})^{-(K/2+a_2)-1}\exp\left(-\frac{1}{\sigma_{i2}^2}\left[\frac{1}{2}\sum_{k=1}^{K}(\mu_{jk}-\mu_{j})^2+b_2\right]\right) \nonumber
	\\ \implies \sigma_{i2}^2\given \bm{t},\mu_{j},\mu_{jk} &\sim IG\left (\frac{K}{2}+a_2, \frac{1}{2} \sum_{k=1}^{K}(\mu_{jk}-\mu_{j})^2+b_2 \right ).
\end{align*}

\section{Metropolis-Hastings ratios} \label{S-sec:MH}
Following \cite{Chipman98} but using $g(\cdot,\cdot)$ rather than $q(\cdot,\cdot)$, the Metropolis-Hastings ratio is given by:
\begin{align}
	\underbrace{\frac{g(T^*,T^{(\ell)})}{g(T^{(\ell)},T^*)}}_\text{transition ratio}\cdot\underbrace{\frac{p(\bm{Y}\given \bm{X},T^*)}{p(\bm{Y}\given \bm{X},T^{(\ell)})}}_\text{likelihood ratio}\cdot\underbrace{\frac{P(T^*)}{P(T^{(\ell)})}}_\text{prior ratio}
\end{align}
Depending on the proposal distribution and type of move in the tree-generating process, the functions $p_\text{SPLIT}(\eta,T)$ and $p_\text{RULE}(\rho\given \eta,T)$ may appear in the transition ratio, and always appear in the prior ratio, potentially leading to cancellations. We use uniform proposal distributions over all variables $J$ and observed values $n_J$, without restricting to splits that result in non-empty nodes for ease of implementation. Uniform proposals are not drawn from our Dirichlet-Multinomial priors, and so this cancellation does not occur for partial pooling. We assume randomly drawing nodes is independent of randomly drawing splitting rules. Each move has a probability of $1/4$ of being chosen, except when starting from a single root node, when only Grow can be chosen, which necessitates a correction by a factor of 4.

Likelihood ratios are evaluated using closed form expressions (Equation 11 in \cite{Chipman98}), and so we focus on the transition and prior ratios in this section. We will leave $p_\text{SPLIT}$ unspecified.

\subsection{Grow step}
\begin{enumerate} 
	\item Select a terminal node, $\eta$, with probability $1/B$ where $B$ is the number of current terminal nodes
	\item Split $\eta$ into two new terminal nodes $\eta_1, \eta_2$ according to rule $\rho$. 
\end{enumerate}
Transition kernel: \begin{align*}
	g(T^{(\ell)},T^*) = P(T^*\given T^{(\ell)}) &= P(\text{node }\eta, \text{ rule }\rho, \text{ grow})
	\\ &= P(\text{node }\eta, \text{ rule }\rho\given \text{grow}) P(\text{grow})
	\\ &= P(\text{node }\eta\given \text{grow}) P(\text{ rule }\rho\given \text{grow}) P(\text{grow})
	\\ &= \frac{1}{B}\cdot \frac{1}{J}\cdot \frac{1}{n_J}\cdot\frac{1}{4} \end{align*}
The reverse proposal distribution, $g(T^*,T^{(\ell)})$, is simply a Prune step with $B_*$ parents of terminal nodes (see the next section).

Prior ratio: $T^*$ only differs from $T^{(\ell)}$ by splitting node $\eta$ into two children, so the ratio only differs by the probability of this new sub-structure.
\begin{align*}
	\frac{P(T^*)}{P(T^{(\ell)})} &= \frac{P(\text{node }\eta \text{ split into }\eta_1, \eta_2)}{P(\text{node }\eta \text{ remains terminal})}
	\\ &= \frac{(1-p_\text{SPLIT}(\eta_1))(1-p_\text{SPLIT}(\eta_2))p_\text{SPLIT}(\eta)p_\text{RULE}(\rho\given \eta,T^{(\ell)})}{1-p_\text{SPLIT}(\eta)}
	\\ &= {\footnotesize\selectfont \frac{(1-p_\text{SPLIT}(\eta_1))(1-p_\text{SPLIT}(\eta_2))p_\text{SPLIT}(\eta)P(\text{variable }j\text{ split}\given \eta,T_k)P(c ^{th}\text{ quantile, split variable }x_j) }{1-p_\text{SPLIT}(\eta)} }
	\\ &= \frac{(1-p_\text{SPLIT}(\eta_1))(1-p_\text{SPLIT}(\eta_2))p_\text{SPLIT}(\eta)P(\text{variable }j\text{ split}\given \eta,T_k)s_j\cdot q_{jc}}{1-p_\text{SPLIT}(\eta)}
\end{align*}

Combining the proposal and prior ratios:
\begin{align*}
	\frac{g(T^*,T^{(\ell)})}{g(T^{(\ell)},T^*)}\frac{P(T^*)}{P(T^{(\ell)})} &= \frac{BJn_J(1-p_\text{SPLIT}(\eta_1))(1-p_\text{SPLIT}(\eta_2))p_\text{SPLIT}(\eta)s_j\cdot q_{jc}}{B_*(1-p_\text{SPLIT}(\eta))}
\end{align*}

\subsection{Prune step}
\begin{enumerate}
	\item Select a parent, $\eta$, of two terminal child nodes with probability $1/B_*$ where $B_*$ is the number of parents of two terminal nodes
	\item Make node $\eta$  a terminal node.
\end{enumerate}
Transition kernel: \begin{align*}
	g(T^{(\ell)},T^*) = P(T^*\given T^{(\ell)}) &= P(\text{node }\eta, \text{ prune }p)
	\\ &= P(\text{node }\eta\given \text{prune}) P(\text{prune})
	\\ &= \frac{1}{B_*}\cdot\frac{1}{4}
	\\ &= \frac{1}{4B_*}.
\end{align*}
The opposite step is just a Grow step, starting from $B$ terminal nodes. Prior ratio:
\begin{align*}
	\frac{P(T^*)}{P(T^{(\ell)})} &= \frac{P(\text{node }\eta\text{ terminal})}{P(\text{node }\eta \text{ split into }\eta_1, \eta_2)}
	\\ &= \frac{1-p_\text{SPLIT}(\eta)}{(1-p_\text{SPLIT}(\eta_1))(1-p_\text{SPLIT}(\eta_2))p_\text{SPLIT}(\eta)p_\text{RULE}(\rho\given \eta,T^{(\ell)})}
	\\ &= \frac{1-p_\text{SPLIT}(\eta)}{(1-p_\text{SPLIT}(\eta_1))(1-p_\text{SPLIT}(\eta_2))p_\text{SPLIT}(\eta)s_j\cdot  q_{jc}} \\
	\\ \frac{g(T^*,T^{(\ell)})}{g(T^{(\ell)},T^*)}\frac{P(T^*)}{P(T^{(\ell)})} &= \frac{b_*(1-p_\text{SPLIT}(\eta))}{BJn_J(1-p_\text{SPLIT}(\eta_1))(1-p_\text{SPLIT}(\eta_2))p_\text{SPLIT}(\eta)s_j \cdot q_{jc}}.
\end{align*}
\subsection{Change step}
\begin{enumerate}
	\item Randomly pick an internal node, $\eta$, with probability $1/B_I$ where $B_I$ is the number of current internal nodes
	\item Assign a new rule $\rho$ to the chosen node $\eta$.
\end{enumerate}
Transition kernel: \begin{align*}
	g(T^{(\ell)},T^*) = P(T^*\given T^{(\ell)}) &= P(\text{node }\eta, \text{ rule }\rho, \text{ change})
	\\ &= P(\text{node }\eta, \text{ rule }\rho\given \text{change}) P(\text{change})
	\\ &= P(\text{node }\eta\given \text{change}) P(\text{ rule }\rho\given \text{change}) P(\text{change})
	\\ &= \frac{1}{B_I}\cdot\frac{1}{J}\cdot\frac{1}{n_J}\cdot\frac{1}{4}.
	\\  &= g(T^*,T^{(\ell)}).
\end{align*} 

Prior ratio: provided the only part of the tree $p_\text{SPLIT}$ depends on is node depth, the splitting probabilities are the same in both $T^*$ and $T^{(\ell)}$, since exactly the same internal node is split and the same child nodes are not split, only the rules differ. Therefore
\begin{align*}
	\frac{P(T^*)}{P(T^{(\ell)})} &= \frac{P(\text{node }\eta \text{ split into }\eta_1, \eta_2\text{ with rule 1})}{P(\text{node }\eta \text{ split into }\eta_1, \eta_2\text{ with rule 2})}
	\\ &= \frac{(1-p_\text{SPLIT}(\eta_1))(1-p_\text{SPLIT}(\eta_2))p_\text{SPLIT}(\eta)p_\text{RULE}(\rho_1\given \eta,T^*)}{(1-p_\text{SPLIT}(\eta_1))(1-p_\text{SPLIT}(\eta_2))p_\text{SPLIT}(\eta)p_\text{RULE}(\rho_2\given \eta,T^{(\ell)})}
	\\ &=
	\frac{p_\text{RULE}(\rho_1\given \eta,T^*)}{p_\text{RULE}(\rho_2\given \eta,T^{(\ell)})}
	\\ \implies
	\frac{g(T^*,T^{(\ell)})}{g(T^{(\ell)},T^*)}\frac{P(T^*)}{P(T^{(\ell)})} &=  \frac{\frac{1}{B_I}\cdot\frac{1}{J}\cdot\frac{1}{n_J}\cdot\frac{1}{4}}{\frac{1}{B_I}\cdot\frac{1}{J}\cdot\frac{1}{n_J}\cdot\frac{1}{4}}\frac{s_{j_1}\cdot q_{j_1c_1}}{s_{j_2}\cdot q_{j_2c_2}} \\&= \frac{s_{j_1} q_{j_1c_1}}{s_{j_2}  q_{j_2c_2}}.
\end{align*}
\subsection{Swap step}
\begin{enumerate}
	\item Select a pair of internal parent and child nodes and swap their splitting rules.
	\item Also swap the other child's rule with the parent if both child nodes have the same original rule.
\end{enumerate}
Transition kernel: \begin{align*}
	g(T^{(\ell)},T^*) = g(T^*,T^{(\ell)}),
\end{align*} 
since the tree structure does not change in this move (only the ordering of splits).
Like for the change step, the prior ratio reduces to the ratio of splitting rule probabilities since the trees have exactly the same structure. There are two cases to consider, based on whether both child nodes have the same rule, which leads to an extra term in both numerator and denominator. Let $\eta_1$ represent the left child node, and since each node considered here is internal, each has its own $p_\text{RULE}$.

Case 1: child nodes with different rules
\begin{align*}
	\frac{P(T^*)}{P(T^{(\ell)})} &= {\footnotesize\selectfont \frac{P(\text{node }\eta \text{ split into }\eta_1, \eta_2\text{ with rule }\rho_1^*)P(\text{node }\eta_1 \text{ split with rule }\rho^*)\cancel{P(\text{node }\eta_2 \text{ split with rule }\rho_2^*)}}{P(\text{node }\eta \text{ split into }\eta_1, \eta_2\text{ with rule }\rho^*)P(\text{node }\eta_1 \text{ split with rule }\rho_1^*)\cancel{P(\text{node }\eta_2 \text{ split with rule }\rho_2^*)}} }
	\\ &= \frac{\cancel{p_\text{SPLIT}(\eta)}p_\text{RULE}(\rho_1^*\given \eta,T^*)\cancel{p_\text{SPLIT}(\eta_1)}p_\text{RULE}(\rho^*\given \eta_1,T^*)}{\cancel{p_\text{SPLIT}(\eta)}p_\text{RULE}(\rho^*\given \eta,T^{(\ell)})\cancel{p_\text{SPLIT}(\eta_1)}p_\text{RULE}(\rho_1^*\given \eta_1,T^{(\ell)})}
	\\ &=
	\frac{p_\text{RULE}(\rho_1^*\given \eta,T^*)p_\text{RULE}(\rho^*\given \eta_1,T^*)}{p_\text{RULE}(\rho^*\given \eta,T^{(\ell)})p_\text{RULE}(\rho_1^*\given \eta_1,T^{(\ell)})}
	\\ &=1,
\end{align*}
since the same draw from $p_\text{RULE}$ occurs in both the numerator and denominator, as the dependence on $\eta,T$ is not important for this model.

Case 2: child nodes with the same rule
\begin{align*}
	\frac{P(T^*)}{P(T^{(\ell)})} &= {\footnotesize\selectfont \frac{P(\text{node }\eta \text{ split into }\eta_1, \eta_2\text{ with rule }\rho_1^*)P(\text{node }\eta_1 \text{ split with rule }\rho^*)P(\text{node }\eta_2 \text{ split with rule }\rho^*)}{P(\text{node }\eta \text{ split into }\eta_1, \eta_2\text{ with rule }\rho^*)P(\text{node }\eta_1 \text{ split with rule }\rho_1^*)P(\text{node }\eta_2 \text{ split with rule }\rho_1^*)} }
	\\&=	\frac{p_\text{RULE}(\rho_1^*\given \eta,T^*)p_\text{RULE}(\rho^*\given \eta_1,T^*)p_\text{RULE}(\rho^*\given \eta_2,T^*)}{p_\text{RULE}(\rho^*\given \eta,T^{(\ell)})p_\text{RULE}(\rho_1^*\given \eta_1,T^{(\ell)})p_\text{RULE}(\rho_1^*\given \eta_2,T^{(\ell)})}.
\end{align*}
The transition ratios always cancel regardless of $p_\text{RULE}$, but when the tree prior uses the Dirichlet-Multinomial models, the prior ratio only cancels in case 1. For case 2 (where likewise the dependence on the particular node and tree is not important for this model):
\begin{align*}
	\frac{P(T^*)}{P(T^{(\ell)})} &=	\frac{p_\text{RULE}(\rho^*\given \eta,T^{(\ell)})}{p_\text{RULE}(\rho_1^*\given \eta,T^{(\ell)})} 
	\\ &= \frac{s_{j_1} q_{j_1c_1}}{s_{j_2}  q_{j_2c_2}}.
\end{align*}
\section{Simulation study}

\begin{figure}
	\centering
	\includegraphics[width=\linewidth]{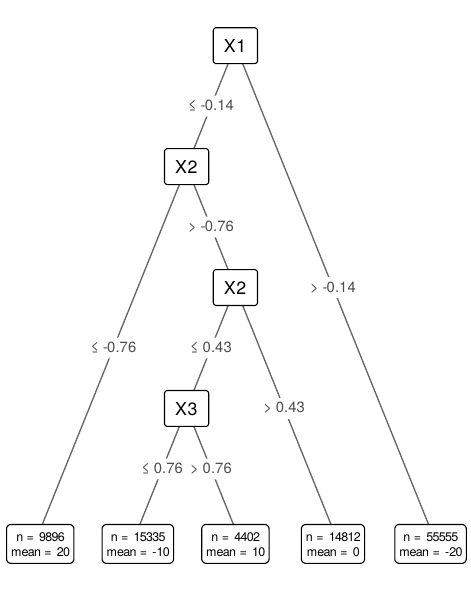}
	\caption{True tree for simulation study.}
	\label{S-fig:example_tree}
\end{figure}

\begin{figure}
	\centering
	\includegraphics[width=\linewidth]{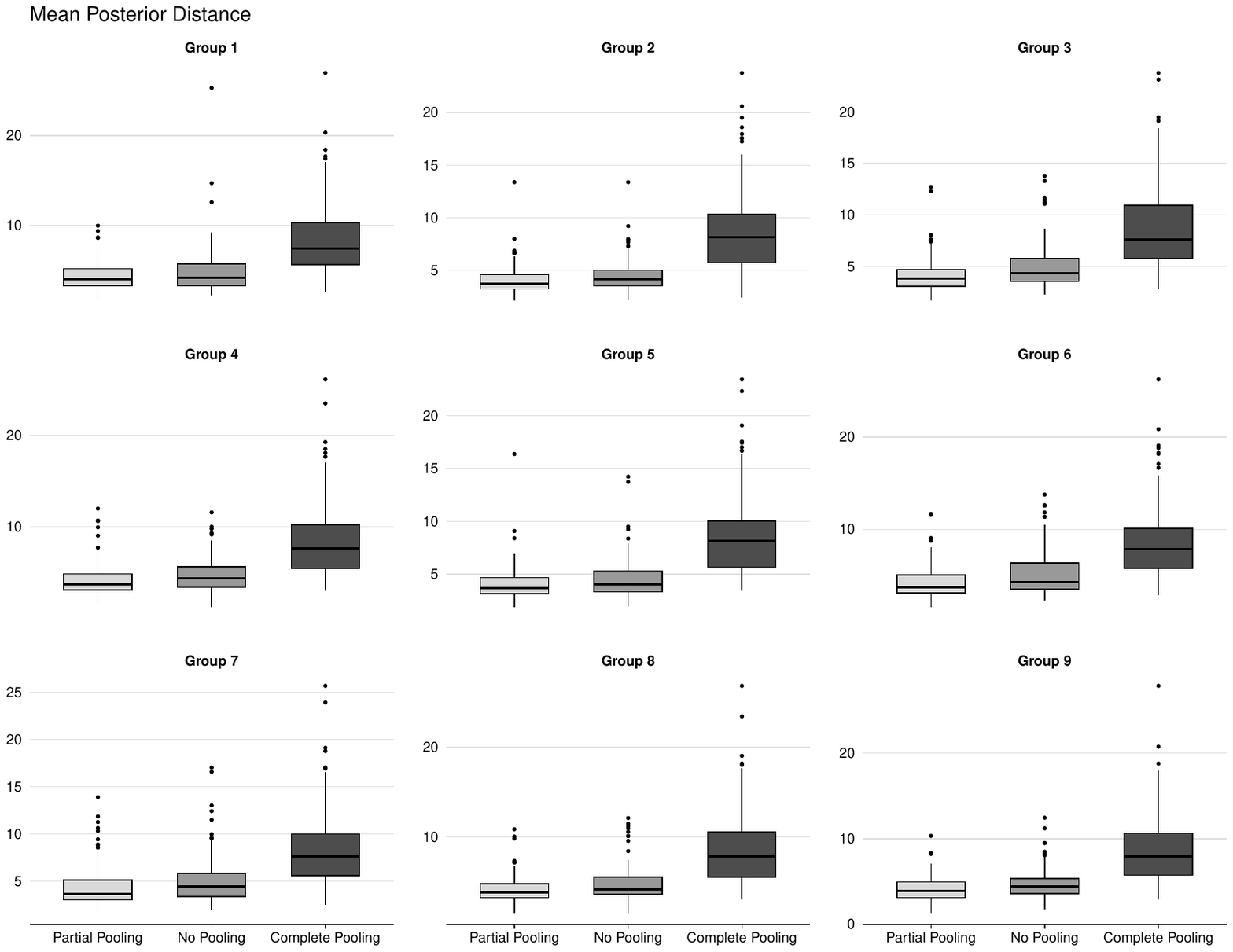}
	\caption{Simulation study mean distance from the true tree in each group, replicated 100 times at random seeds.}
	\label{S-fig:group_MPD}
\end{figure}

\begin{figure}
	\centering
	\includegraphics[width=0.8\linewidth]{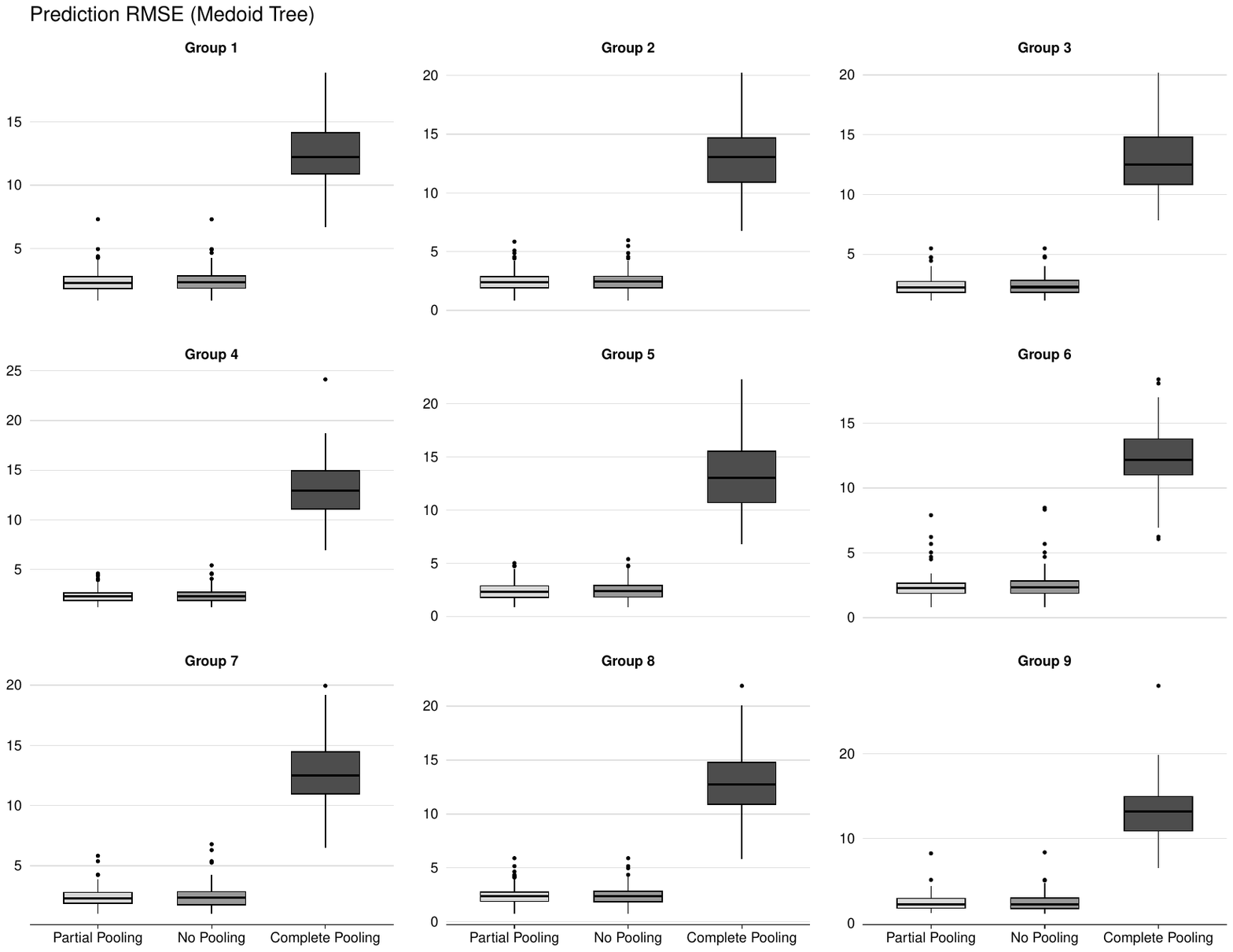}
	\caption{Simulation study prediction error on test set using medoid tree in each group, calculated as Root Mean Square Error (RMSE), replicated 100 times at random seeds.}
	\label{S-fig:group_RMSE}
\end{figure}

An example true tree from which data were generated for the simulation study is shown in Figure \ref{S-fig:example_tree}. Boxplots for each of the 9 groups of mean posterior distance and out-of-sample test set RMSE are shown in Figures \ref{S-fig:group_MPD} and \ref{S-fig:group_RMSE}.

\section{Wildfire data}

90\% interval coverage for each method is reported in Table \ref{S-tab:coverage}. Coverage for pooling models is based on Bayesian posterior quantiles. RPART estimates are from bootstrapping, and multilevel linear (group-specific intercept for ecoregion) is based on profile likelihood calculations.

\begin{table}[htbp]
	\centering
	\caption{Coverage by ecoregion and pooling method. Green indicates values closest to the nominal 90\% target (smallest $\given {\rm coverage}-0.90\given $ overall), yellow indicates the largest deviation from 90\% overall.}
	\label{S-tab:coverage}
	\small
	\resizebox{\textwidth}{!}{%
		\begin{tabular}{lS[table-format=1.4]S[table-format=1.4]S[table-format=1.4]S[table-format=1.4]S[table-format=1.4]}
			\toprule
			Ecoregion & {\textbf{Partial Pooling}} & {\textbf{No Pooling}} & {\textbf{Complete Pooling}} & {\textbf{rpart}} & {\textbf{Multilevel linear}} \\
			\midrule
			Cold Deserts & \cellcolor{cbHigh!91!cbLow}{0.9092} & \cellcolor{cbHigh!81!cbLow}{0.9178} & \cellcolor{cbHigh!94!cbLow}{0.9065} & \cellcolor{cbHigh!66!cbLow}{0.9300} & \cellcolor{cbHigh!77!cbLow}{0.9211} \\
			Marine West Coast Forest & \cellcolor{cbHigh!48!cbLow}{0.9457} & \cellcolor{cbHigh!58!cbLow}{0.9374} & \cellcolor{cbHigh!10!cbLow}{0.9777} & \cellcolor{cbHigh!47!cbLow}{0.9466} & \cellcolor{cbHigh!0!cbLow}{0.9861} \\
			Mediterranean California & \cellcolor{cbHigh!98!cbLow}{0.9032} & \cellcolor{cbHigh!83!cbLow}{0.9162} & \cellcolor{cbHigh!30!cbLow}{0.8396} & \cellcolor{cbHigh!56!cbLow}{0.9386} & \cellcolor{cbHigh!35!cbLow}{0.8438} \\
			S. Central Semiarid Prairies & \cellcolor{cbHigh!82!cbLow}{0.9166} & \cellcolor{cbHigh!74!cbLow}{0.9236} & \cellcolor{cbHigh!80!cbLow}{0.9187} & \cellcolor{cbHigh!68!cbLow}{0.9283} & \cellcolor{cbHigh!73!cbLow}{0.9244} \\
			Upper Gila Mountains & \cellcolor{cbHigh!79!cbLow}{0.9192} & \cellcolor{cbHigh!81!cbLow}{0.9179} & \cellcolor{cbHigh!70!cbLow}{0.8730} & \cellcolor{cbHigh!69!cbLow}{0.9275} & \cellcolor{cbHigh!99!cbLow}{0.8972} \\
			Warm Deserts & \cellcolor{cbHigh!77!cbLow}{0.9207} & \cellcolor{cbHigh!69!cbLow}{0.9274} & \cellcolor{cbHigh!69!cbLow}{0.9279} & \cellcolor{cbHigh!60!cbLow}{0.9356} & \cellcolor{cbHigh!72!cbLow}{0.9252} \\
			W-Central Semiarid Prairies & \cellcolor{cbHigh!79!cbLow}{0.9194} & \cellcolor{cbHigh!76!cbLow}{0.9220} & \cellcolor{cbHigh!71!cbLow}{0.9262} & \cellcolor{cbHigh!65!cbLow}{0.9310} & \cellcolor{cbHigh!60!cbLow}{0.9352} \\
			Western Cordillera & \cellcolor{cbHigh!77!cbLow}{0.9208} & \cellcolor{cbHigh!77!cbLow}{0.9213} & \cellcolor{cbHigh!76!cbLow}{0.9219} & \cellcolor{cbHigh!64!cbLow}{0.9322} & \cellcolor{cbHigh!62!cbLow}{0.9336} \\
			W. Sierra Madre Piedmont & \cellcolor{cbHigh!99!cbLow}{0.9027} & \cellcolor{cbHigh!100!cbLow}{0.9016} & \cellcolor{cbHigh!8!cbLow}{0.8208} & \cellcolor{cbHigh!91!cbLow}{0.9093} & \cellcolor{cbHigh!51!cbLow}{0.8568} \\
			\bottomrule
		\end{tabular}%
	}
\end{table}

\section{Medoid trees for larger ecoregions}
\clearpage

\includepdf[
pages=1, fitpaper=true,
pagecommand={\thispagestyle{empty}%
	\captionof{figure}{Medoid tree for Cold Deserts.}%
	\label{S-fig:supp_cold_deserts}}
]{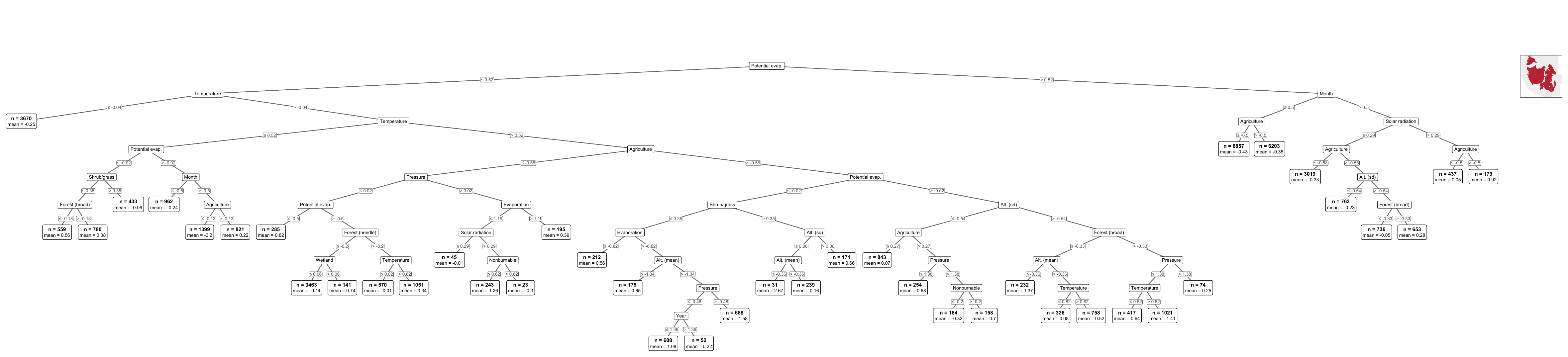}

\includepdf[
pages=2, fitpaper=true,
pagecommand={\thispagestyle{empty}%
	\captionof{figure}{Medoid trees for Mediterranean California and Warm Deserts.}%
	\label{S-fig:supp_medcal_warm}}
]{medoid_trees_supplementary_material.pdf}

\includepdf[
pages=3, fitpaper=true,
pagecommand={\thispagestyle{empty}%
	\captionof{figure}{Medoid tree for West-Central Semiarid Prairies.}%
	\label{S-fig:supp_westcentral}}
]{medoid_trees_supplementary_material.pdf}

\includepdf[
pages=4, fitpaper=true,
pagecommand={\thispagestyle{empty}%
	\captionof{figure}{Medoid tree for Western Cordillera.}%
	\label{S-fig:supp_cordillera}}
]{medoid_trees_supplementary_material.pdf}

\bibliography{references}

\end{document}